\documentclass[12pt,a4paper]{article}
\usepackage{jheppub}
\usepackage{amsmath}
\usepackage{amsfonts}
\usepackage{mathtools}
\usepackage{amssymb}
\usepackage{caption}
\usepackage{subcaption}

\title{Thermalization of Green functions and
quasinormal modes}
\author{Justin R. David, Surbhi Khetrapal}
\affiliation{Centre for High Energy Physics, Indian Institute of Science,\\
C. V. Raman Avenue, Bangalore 560012, India.}
\emailAdd{justin, surbhi@cts.iisc.ernet.in}
\abstract{ We  develop a new method to 
study the thermalization of  time dependent retarded Green function in conformal field theories
holographically dual to thin shell $AdS$ Vaidya space times.  The method relies on
using the information  of all  time derivatives of the 
Green function at the shell  and then evolving it for later times.
The time derivatives of the Green function  at the shell is given in terms of a recursion formula. 
 Using this method 
we  obtain analytic results for short time  thermalization of the Green function. 
We  show that the late time behaviour of  the Green function 
is  determined by the first quasinormal mode. 
We then implement the method  numerically. As applications of this method
we study the thermalization of  the retarded time dependent
 Green function  corresponding to  a minimally coupled 
scalar  in  the  $AdS_3$  and $AdS_5$  thin Vaidya shells. 
We see that as expected the late time behaviour is 
determined by the first quasinormal mode. We  apply the method to study the late time behaviour
of the shear vector mode  in $AdS_5$ Vaidya shell. At small momentum the 
corresponding time dependent Green function is expected to relax to equilibrium
 by the shear hydrodynamic mode. Using this we obtain   
the universal  ratio of the shear viscosity to entropy density from a  time dependent process. 
}

\begin{document}
\maketitle
\flushbottom

\section{Introduction}

The question of how a highly excited state in a quantum system  relaxes to equilibrium 
or how a quantum system relaxes when one of the parameters describing its 
Hamiltonian is quenched is of phenomenological interest. 
The relaxation of 
the state produced initially by high energy nuclear collisions from  a highly excited state to quark-gluon plasma is an  important question in the RHIC experiments \cite{Jacak:2012dx}. 
 Similarly  the question of quenching of a quantum system can be studied experimentally  in 
 cold atoms where the coupling of an interacting system
  can be tuned to almost any value    and in 
 short times scales \cite{2002Natur.419...51G,PhysRevLett.95.190405,2006Natur.443..312S,Kinoshita}.
 There have been various approaches,  both analytical  and numerical,  developed 
 to study these questions for a variety of quantum systems. 
When the question of thermalization is asked in quantum field theories which admit a 
holographic dual, the 
gauge/gravity correspondence links the question of approach to equilibrium in 
the field theory
 to the formation of a black hole in the bulk.  See 
 \cite{KalyanaRama:1999zj,Danielsson:1999zt,Giddings:1999zu,Danielsson:1999fa,Giddings:2001ii} 
 for  early  studies pursuing this idea.  
 
 More recently motivated by success of the gauge/gravity duality to 
 describe near equilibrium physics and hydrodynamic behaviour 
 in strongly coupled field theories this question has received 
 renewed attention which has resulted in more 
 quantitative understanding 
 \cite{Lin:2008rw,Chesler:2008hg,Bhattacharyya:2009uu,Chesler:2009cy,Beuf:2009cx,Heller:2011ju,Heller:2012je}. 
 In \cite{Bhattacharyya:2009uu} the excited state in the quantum field theory was created  by a translational 
 invariant perturbation along the boundary
  of  a minimally coupled scalar field .  This perturbation lasted for a short duration of time. 
 It was shown by solving the bulk 
 equations  that  for a small amplitude of the perturbation,  the metric outside the 
 in-falling shell of matter is that of a black brane at the leading order. 
 This result encouraged subsequent authors to model the collapse to a black hole  by a homogenous 
 in-falling  shell of matter 
 \cite{AbajoArrastia:2010yt,Albash:2010mv,Ebrahim:2010ra,Balasubramanian:2011ur,Aparicio:2011zy,Balasubramanian:2012tu,Callebaut:2014tva}.

\begin{figure}
\centering
\includegraphics[scale=0.5]{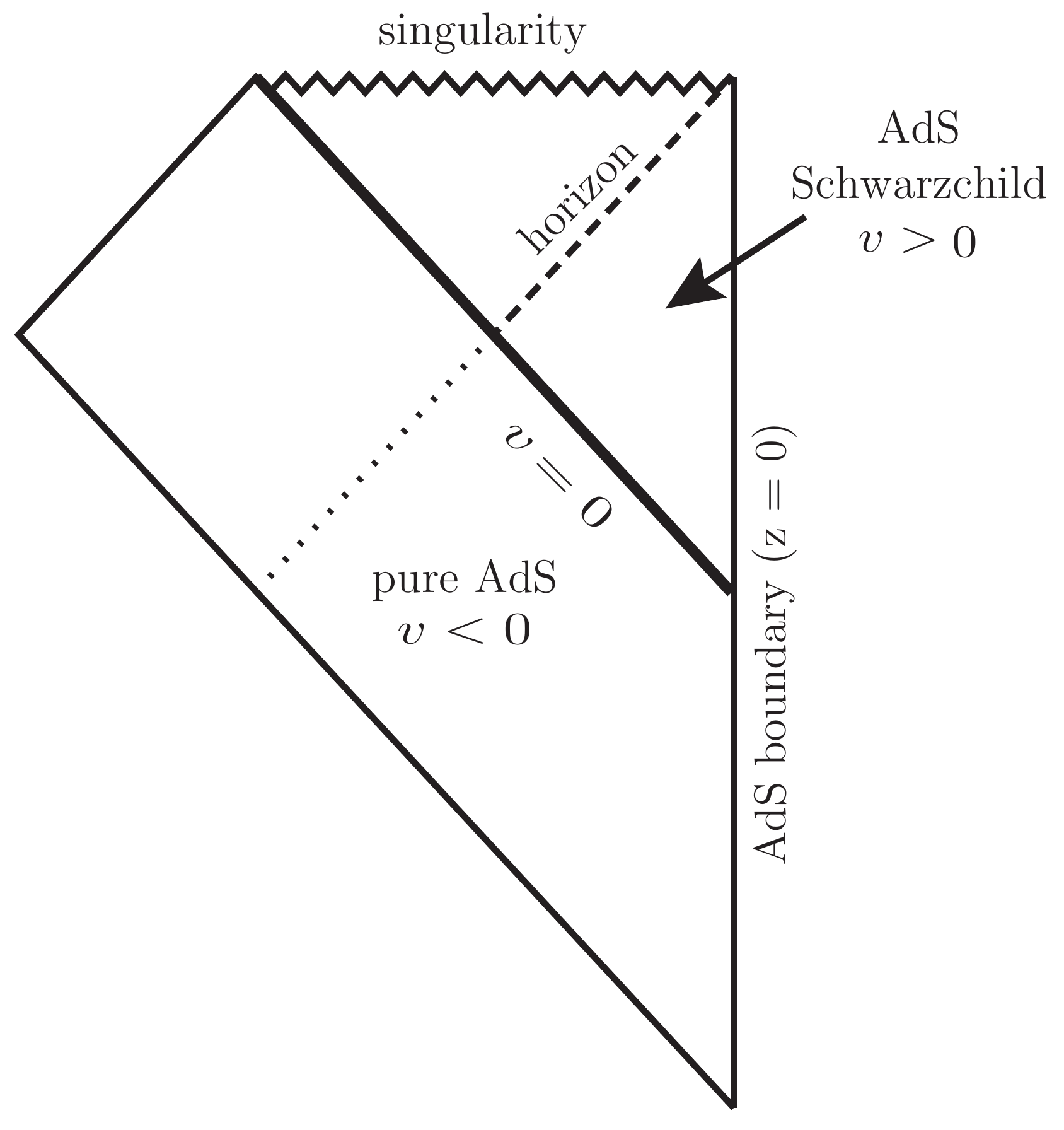}
\caption{The Penrose diagram of collapse in $AdS$ Vaidya.}
\label{diag}
\end{figure}

 For definiteness we consider  the thin shell model of collapse  \cite{Balasubramanian:2011ur}
 given by the following metric
 \begin{eqnarray}\label{vaidya1}
 ds^2 &=& \frac{1}{z^2}  \left[ - ( 1  - \theta(v)   z^d ) dv^2  -2 dz dv  + d {\bf x}^2 \right], \\ \nonumber
 \theta(v)  &=&\left\{   \begin{array}{l}   0 ,  \qquad  \hbox{for} \quad v< 0  , \\
  1, \qquad \hbox{for} \quad v\geq 0 .
 \end{array}\right.
 \end{eqnarray}
 Here $z$ refers to the radial co-ordinate, the boundary is at $z=0$. ${\bf x} = x^1, \cdots x^{d-1}$ are the
 spatial co-ordinates at the boundary. 
 The metric for $v < 0$ can be seen to be that of $AdS_{d+1}$ using the following 
 co-ordinate transformation
 \begin{equation}
 v =  t  - z.
 \end{equation}
 While the metric for $v>0$ reduces to that of the black brane in $AdS_{d+1}$ under the 
 co-ordinate transformation 
 \begin{equation} \label{difftrans}
 dv = dt -  \frac{dz}{ 1- z^d}.
 \end{equation}
 From these co-ordinate transformations, we see that $v$ coincides with  time $t$  at the boundary.  We have chosen to work  with  units
 in which the radius of $AdS_{d+1}$ is unity.  The  radius of the horizon is also unity. The Penrose diagram of the collapse is given  in figure \ref{diag}.
 
 In \cite{Balasubramanian:2011ur} the  study of how probes such as 
 two point functions,  Wilson loops and the entanglement entropy \cite{Hubeny:2007xt}
  thermalize in the thin shell Vaidya collapse  was initiated. 
 This  study was mainly confined to the saddle point approximation of the  probes.  
 The behaviour of the probes were characterized in terms of  their minimal  geometric 
 volume. In \cite{Balasubramanian:2012tu} the analysis was extended to study the 
 two point function of operators dual to a minimally 
 coupled massive scalar beyond the geodesic approximation for the case of the 
 $AdS_3$ Vaidya shell.    
 The  retarded Green function  $G_R(t_2, t_1;k )$ 
 with  time $t_1<0, t_2>0$, before and after the collapse of the shell  
  was evaluated numerically.  Translational invariance in the spatial 
 direction  of the collapsing shell (\ref{vaidya1})
 enabled the  characterization of the Green function in the 
 Fourier   $k$ space corresponding to the spatial directions. 
 The analysis 
 was numerical and it  showed that the  relaxation of the Green function 
is determined by the first quasinormal mode. In  \cite{Callebaut:2014tva}  the study of thermalizing 
Green functions was extended to fermions  
in the $AdS_4$ Vaidya shell. The analysis was again done  numerically. Some analytic properties of the 
time dependent scalar Green functions was studied in \cite{Keranen:2014lna}.

In this paper we develop a new method to evaluate the retarded two point function
$G_R(t_2, t_1;k )$  of an operator in the dual theory corresponding to
 the  collapsing Vaidya $AdS$ shell in the bulk.    The method is general and can be implemented in arbitrary dimensions and for arbitrary types of fields in the bulk. 
The  method relies on performing the matching of the wave functions  corresponding to the dual fields  before and after the shell term by 
term in the expansion  of the radial co-ordinate $z$ of (\ref{vaidya1}). We will show that this enables the determination of all the time derivatives of   $G_R$ just after the collapse, $v=0^+$ of the 
shell.  It is then possible to evolve the Green function to an arbitrary future 
time $t_2$.  On implementing this method  we see that to obtain information of 
more and more higher derivatives of the Green function one needs to perform the matching 
of the wave functions of the bulk fields 
to higher powers in the radial coordinate $z$. This implies that one needs the information of the 
wave functions of these fields  
closer to the horizon to obtain long time behaviour of the Green function. 
It also implies that short time behaviour of the two point function after the collapse
 can be determined analytically 
from the near boundary information of the wave functions.  
Using the fact that the Green function at long time  is determined by the 
near horizon behaviour of the wave functions we show that the 
relaxation of the Green function to equilibrium is determined by the 
first quasi-normal mode of the dual bulk field corresponding to the operator of interest in the black hole background. 

We  implement the method numerically and re-visit the case of the minimally coupled
massive scalar in the  $AdS_3$ Vaidya shell. We reproduce the results of 
\cite{Balasubramanian:2012tu}. We then study the 
case of the minimally coupled massless scalar in $AdS_5$ Vaidya shell.
The  Green function 
corresponding to this scalar is  the retarded two point function of the 
spin $2$ part of the stress tensor.  We show that this Green function relaxes by the 
first quasi-normal mode which was determined numerically in \cite{Nunez:2003eq}. 
We then study the vector fluctuations of the metric and obtain the
Green function of  the vector part of the stress  tensor. 
For small momentum $k$, it is known \cite{Policastro:2002se} that this mode admits a hydrodynamic quasi-normal mode
which obeys  the 
dispersion relation given by 
\begin{equation}
\omega  = - i \frac{\eta}{Ts} k^2,
\end{equation}
where $\eta$ is the shear viscosity, $s$ the entropy density and $T$ the temperature 
of the fluid.  We show the time dependent Green  function corresponding to the 
vector fluctuations of the metric relaxes to equilibrium at small $k$ by the hydrodynamic mode. 
Using this we determine the universal ratio of shear viscosity to entropy density from a time dependent 
process.

This paper is organized as follows. 
In the next section we detail the new method developed in this paper to evaluate the 
retarded Green function in collapsing $AdS$ Vaidya thin shell backgrounds. We see that 
the method results in  a recursion formula for the derivatives of the Green function 
just after the collapse of the shell. In section 3 we apply the method to obtain the 
Green function of the operator dual to the minimally coupled scalar in 
$AdS_3$ Vaidya shell. 
In section 4 we show that the long 
time behaviour of the Green function is determined by the first quasi-normal mode. 
This is first done for the case of $AdS_3$ Vaidya for which wave functions before and
after the collapse of the shell are known exactly. Then the argument is extended 
in general for any Green function. 
In section 5 we turn to the case of $AdS_5$ Vaidya. We first study the 
thermalization of 
the shear correlator of the  stress tensor
by solving the minimally coupled scalar in $AdS_5$ Vaidya. 
We then examine the vector perturbations of the metric to evaluate the 
two point function of the spin-1 part of the stress tensor and show that it relaxes 
at small momentum by the shear hydrodynamic mode. 
Section 6 contains the conclusions.  Appendices A to D deal with technical details required 
for the analysis in the paper. Appendix E describes the Mathematica notebooks  
which enable the evaluation of  the Green functions numerically using the recursive 
method developed in this paper.

\section{Recursion method for time dependent Green functions} \label{sec-gen}

In this section we will outline the general method to obtain the retarded Green function in the thin shell 
Vaidya $AdS$ geometry. 
The method  is general, and can be applied to any field in the $AdS_{d+1}$ Vaidya geometry given in (\ref{vaidya1}).  
For definiteness let us focus on  the minimally coupled scalar of mass $m$. 
The differential equation  obeyed by $\phi$  is given by 
\begin{equation}\label{mincoup}
 h(v, z) \partial_z^2 \phi + \left( \frac{1}{z} \partial_v \phi - 2 \partial_v \partial_z \phi \right) 
 + \left( \partial_ z h  - \frac{h(v, z) }{z} \right) \partial_z \phi
 - \left( \frac{m^2}{z^2}  + k^2 \right) \phi =0,
\end{equation}
where
\begin{equation}
 h(v, z) = 1 - \theta(v) z^d,
\end{equation}
and $k$ is the Fourier conjugate of direction $x^{d-1}$.

The solution $\phi^{AdS}$ in the region $v<0$ corresponding to 
before the formation of the black hole admits a  
closed form in terms of Bessel functions. The analytical solution will be explicitly 
discussed in the examples we will consider subsequently.   The solution has the form
\begin{eqnarray} \label{adssol}
\phi^{AdS}(v,k,z;t_1) &=& z^{\Delta_+} J(v-t_1,k,z),  
\end{eqnarray}
where $\Delta_+>  \Delta_-$ are the two solutions of the  equation
\begin{equation}
 \Delta (\Delta - d) = m^2 .
\end{equation}
The solution we choose satisfies the boundary condition 
\begin{align}\label{gen-0-0}
\phi^{AdS}(v,k,z;t_1) \xrightarrow{z \to 0} z^{\Delta_-} \delta(v-t_1) + \dots,
\end{align}
with $t_1<0$.  This  is necessary to obtain the retarded Green function.
In (\ref{adssol}) note that due to time translational symmetry for $v<0$, the wave function 
just depends on the combination $v-t_1$. 
In the black hole region $v>0$, the bulk equations of motion usually do not admit a closed form solution. 
But the solution in the frequency $\omega$ and momentum $k$ domain can be constructed in terms of a Frobenius series around the boundary $z=0$. Then the most general solution in the 
time domain $v>0$ can be obtained  by taking Fourier transform of the two independent 
solutions obtained by the Frobenius  method  with respect to the  frequency, $\omega$. 
We write this as  
\begin{align}\label{gen-0-1}
\phi^{BH}(v,k,z)=\int_{-\infty}^\infty d\omega e^{-i\omega v} \sum_{n=0}^\infty \left[ z^{\Delta_+} C(\omega,k)  A_n(\omega,k) z^n 
+ z^{\Delta_-} D(\omega,k) B_n(\omega,k) z^n \right]. 
\end{align}
Note that here we have assumed that the roots of the indicial equation of (\ref{mincoup}), $\Delta_+, \Delta_-$  do not 
differ from each other by an integer. The discussion can be carried out for the case when the roots 
differ by an integer but as we will see that we will only need the less singular solution which falls of as $z^{\Delta_+}$ to construct the 
retarded Green function.  $C(\omega, k)$ and $D(\omega, k) $  in (\ref{gen-0-1}) are functions which must be determined by
continuity 
at $v=0$.  Note that the coefficients of the differential equation (\ref{mincoup}) are discontinuous, but the discontinuity is 
finite across $v=0$, therefore the solution $\phi$ must be continuous at $v=0$. 
Thus we have 
\begin{align} \label{gen-1}
\phi^{AdS}(v=0,k,z;t_1)=\phi^{BH}(v=0,k,z).
\end{align}
We impose  continuity  by equating  each term of the power series in $z$ about the boundary. 
Thus we expand  both sides of (\ref{gen-1})  in powers of $z$ and obtain  the equation
\begin{eqnarray}\label{gen-0-2}
& &  z^{\Delta_+} \sum_{n=0}^\infty  \tilde J_n   z^n + z^{\Delta_-}  \delta( -t_1)   \\ \nonumber
 && \qquad\qquad\qquad = \int_{-\infty}^\infty d\omega \sum_{n=0}^\infty
 \left( z^{\Delta_+} C(\omega)  A_n(\omega ) z^n + z^{\Delta_-} D(\omega) B_n(\omega ) z^n \right).
\end{eqnarray}
We have suppressed the dependence of $\tilde J_n$ on $t_1, k$ and the dependence of $C(\omega) , A_n,$ $ D(\omega), B_n$ on 
$k$ to un-clutter the equations. Equating the coefficients of $z^n$ in the  terms proportional to  $z^{\Delta_+}$ we obtain 
\begin{align}\label{defjn}
 \tilde J_n   =  \int_{-\infty}^\infty d\omega \, C(\omega ) A_n(\omega).
\end{align}
From the examples considered in the paper, it is seen that $A_n(\omega)$ is an $n$-th order polynomial in $\omega$
\footnote{This can be seen using the recursion formula obtained during the construction of the 
Frobenius series solution of (\ref{mincoup}). }. Therefore we write $A_n$ as 
\begin{align} \label{defan}
A_n = \sum_{j=0}^n a_n^j \omega^j. 
\end{align}
Substituting this expansion in (\ref{defjn}) we obtain 
\begin{align} \label{gen-2}
\tilde J_n = \sum_{j=0}^n a^j_n \int_{-\infty}^\infty d\omega \, C(\omega)  \omega^j.
\end{align}
Let us define 
the $j$-th moment of $C(\omega)$ as,
\begin{align}
M_j = \int_{-\infty}^\infty d\omega \, C(\omega) \omega^j.
\end{align} 
Substituting this in equation (\ref{gen-2}), we can rewrite it as, 
\begin{align}
\tilde J_n = \sum_{j=0}^n a_n^j M_j.
\end{align}
This equation can be  inverted to obtain the moments, $M_j$, which contain information about $C(\omega)$. 
Note that  $a_n^j$ are known from Frobenius series solution  $\phi^{BH}$.
Therefore we obtain 
\begin{align}\label{recurm}
M_n & = \frac{1}{a^n_n}\left(\tilde J_n - \sum_{j=0}^{n-1} a_n^j M_j\right); \quad n>0,\\
M_0 & = \frac{\tilde J_0}{a_0^0}.
\end{align}

Knowledge of all the moments  $M_j$ is sufficient to construct the retarded Green function.  To see this  
consider the near boundary  behaviour of the field $\phi$. 
From equations (\ref{gen-0-0}), (\ref{gen-0-1})  and (\ref{gen-0-2}), the near boundary behaviour of the field  for $v>0$ is 
given by 
\begin{align}
\phi^{BH} (v, t_1, k)  =  \int_{-\infty}^\infty d\omega e^{-i\omega v} C(\omega) z^\Delta_+ A_0 +  \cdots +
O( z^{\Delta_-})  .
\end{align}
Together with  the boundary condition (\ref{gen-0-0}), 
 the AdS/CFT recipe  for the retarded Green function  \cite{Son:2002sd}  states that
it is given by 
\begin{align} \label{sum_moments}
G_R(v) =  \int_{-\infty}^\infty d\omega e^{-i\omega v} C(\omega).
\end{align}
Here we have ignored overall proportionality constants  in the Green function 
to simplify the discussion. 
Expanding  the exponential as Taylor series, and using the definition of the moments of $C(\omega)$ we obtain
\begin{align}\label{greenp}
G_R(v) = \sum_{n=0}^\infty \frac{ (-iv)^n M_n}{n!}.
\end{align}
Thus the knowledge of all the moments of  $C(\omega)$ is sufficient to construct the Green function. Note that knowledge of 
$D(\omega)$ in (\ref{gen-0-1}) is not necessary. 
We call this the recursion method to obtain the Green function since each  term in 
(\ref{greenp}) is given recursively 
from the knowledge of the lower moments using (\ref{recurm}). It allows for the construction 
of the the Green function as 
a power series in time for $v>0$.   
The information of $t_1$ is present in $\tilde J_n$ and all terms 
depend on momentum $k$. 
There is another way to view this construction of the retarded Green  function. 
Note that  the $j$-th moment of $C(\omega)$ is  essentially the $j$-th derivative of the Green function, evaluated at $v=0^+$,
\begin{align}
\frac{\partial^j  G_R}{\partial v^j} \Big|_{v=0^+} = \int_{-\infty}^\infty d\omega \, (-i\omega)^j C(\omega) = (-i)^j M_j.
\end{align} 
Thus this method  determines the Green function for $v>0$ from the knowledge of all its time derivatives 
at $v=0^+$. 

We now  make some general properties of this method of determining the retarded Green function. 
Since the method relies on construction of $\phi^{BH}$ using the  Frobenius series we can apply 
it in general to all black hole backgrounds even if closed form solutions do not exist. 
The method can be applied even if the background is only known in terms of a 
power series in $z$ about the boundary for example in finite 
temperature versions of RG flow solutions  \cite{Buchel:2003ah}. 
Note that for short time development of the Green function for $v>0$, knowledge of only a few moments  is needed. 
This implies from (\ref{greenp}) and (\ref{recurm})  we need 
the knowledge of wave functions before and after $v=0$ close to the boundary.  That is we need
the knowledge of the wave functions to a  few  powers of $z$.  
Thus short time development of the Green function can be written down  analytically by obtaining
a few moments.  
However for the long time behaviour of the Green function we need to know a large number of moments. 
Again from (\ref{recurm}), 
this implies we need the information of the wave function for large powers of $z$ which in turn 
implies that we need the behaviour close to the horizon. 
This fits with the general intuition that long time behaviour is controlled by the behaviour near the horizon. 
It is this property which will enable us to prove that the long time behaviour of the Green function 
is determined by the quasinormal mode in section \ref{longt}.  
Finally, it will turn out that  
even moments are real and odd moments are purely imaginary. 
This is because the coefficients $a_n^j$ for odd $j$  are imaginary. 
This is easily seen due to the fact that each power of $\omega$ comes with a factor of $i$. 
Then from (\ref{recurm}) it is easy to see that even moments are real and odd moments are purely imaginary. 
This then ensures that the 
Green function given by (\ref{greenp}) is real. 

\section{Thermalization in $AdS_3$ Vaidya}

In this section we will implement the method developed in section \ref{sec-gen} for the case of 
thin shell Vaidya metric in $AdS_3$.  We  study the thermalization of the retarded Green function of the operator
corresponding to the minimally coupled scalar of mass $m$. 
We will show that using our method we reproduce the results of \cite{Balasubramanian:2012tu}.  
We also determine a few low moments 
analytically to obtain the short time behaviour of the Green function. 
A simplification that occurs for the case of the $AdS_3$ is that the solutions of the minimally coupled scalar in the 
BTZ black hole are known in closed form in terms of hypergeometric functions \cite{Birmingham:2001hc}. 
These solutions will provide the initial starting point  in our argument to demonstrate that the 
long time behaviour of the Green function is determined by the lowest quasi-normal mode. 

The thin-shell $AdS$-Vaidya metric, in $3$ dimensions is given by 
\begin{align}
\begin{split} \label{ads3-3}
ds^2 & = \frac{1}{z^2} [-h(z, v) dv^2 - 2 dv dz + d\phi^2],\\
& h(z, v)  = 1 - \theta(v) z^2.
\end{split}
\end{align}
The co-ordinate $\phi$ parametrizes the spatial direction of the field theory and we assume that it is not 
compact. 
The scalar field equation  in this metric is given by 
\begin{equation} \label{ads3-4}
h \partial^2_z \Phi + \left( \frac{1}{z} \partial_v \Phi - 2 \partial_v \partial_z \Phi \right) + \left( \partial_z h - \frac{h}{z} \right) \partial_z \Phi - \left( \frac{m^2}{z^2}+k^2 \right) \Phi = 0.
\end{equation}

\subsection{Scalar wave functions in $AdS_3$ Vaidya shell}

We  solve the minimally coupled massive scalar equation given in (\ref{ads3-4}). We discuss the 
solutions before and after the collapse of the shell below.  

\vspace{.5cm}
\noindent
{\bf Solution for $v<0$}
\vspace{.5cm}

\noindent
The solution for the scalar field  for $v<0$ is given by \cite{Balasubramanian:2012tu}
\begin{eqnarray} \label{ads3-5}
&& \Phi^{AdS}(v-t_1,k,z) \\
&& \quad = C \frac{\theta(v-t_1) z^{\Delta_+}}{[(v-t_1)^2+2(v-t_1)z]^{\frac{2\nu+1}{4}}} |k|^{\nu+\frac{1}{2}} J_{-\nu-\frac{1}{2}}\left( |k| \sqrt{(v-t_1)^2+2(v-t_1)z} \right),\nonumber\\
&& C = \frac{2^{\frac{1}{2}-\nu} \sqrt{\pi}}{\Gamma(\nu)}.
\end{eqnarray}
Note that this solution is written down  in mixed Fourier space $(t, k)$. 
Here  $\Delta_\pm$ are solutions to  the quadratic equation $\Delta(\Delta-2)=m^2$ and are given by 
\begin{align}
\Delta_\pm = 1 \pm \nu, \quad \nu = \sqrt{1+m^2}.
\end{align}
This solution satisfies the  required boundary condition  discussed in  (\ref{gen-0-0})  which is required to evaluate the 
retarded Green  function,
\begin{equation}\label{bcdelta}
\Phi^{AdS} (v-t_1,k,z) = \delta(v-t_1) z^{\Delta_-} + \dots .
\end{equation}
To show this we first take the  the limit $z\rightarrow 0$ in (\ref{ads3-5})  keeping $v - t_1 \neq 0$. 
We see that there is no term proportional to $ z^{\Delta_-}$. 
Now we take $v\rightarrow t_1$, the wave function then reduces to 
\begin{equation}\label{limdelt}
\Phi^{AdS} (k,v,z)_{z \rightarrow 0 } =  
\frac{\theta(v-t_1) 2 \sqrt{\pi}}{\Gamma(-\nu+\frac{1}{2})
\Gamma(\nu)} \frac{z^{\Delta_+}}{[(v-t_1)^2+2(v-t_1)z]^{\nu+\frac{1}{2}}} .
\end{equation}
This solution certainly diverges  in the limit $v\rightarrow t_1$. 
 All what one needs to show  is that 
the expression in (\ref{limdelt})  is a representation of the delta function  times $z^{\Delta_-}$.  To  demonstrate this  we 
perform the integral over $v$ as follows 
 \begin{equation}
 \begin{split}
\int_{-\infty}^{\infty} dv \,  \Phi^{AdS} (k,v,z)_{z \rightarrow 0}  & =  \frac{ 2 \sqrt{\pi}}{\Gamma(-\nu+\frac{1}{2}) 
 \Gamma(\nu)}z^{\Delta_-} \int_{-\infty}^\infty \frac{d v}{z} 
 \theta(v-t_1) \frac{1}{[(\frac{v-t_1}{z}+1)^2-1]^{\nu+\frac{1}{2}}}\\
 & = z^{\Delta_-} .
 \end{split}
 \end{equation}
 Therefore we conclude  that the solution in (\ref{ads3-5}) satisfies the required boundary condition given in (\ref{bcdelta}).

\vspace{.5cm}
\noindent
{\bf Solution for $v>0$}
\vspace{.5cm}

The strategy to obtain a closed form solution for $v>0$ is as follows. Note that under the transformation 
\begin{equation}\label{vtos}
t = v - \frac{1}{2} \ln \left( \frac{ 1- z }{ 1+ z} \right) ,
\end{equation}
the metric  given in (\ref{ads3-3}) for $v>0$ reduces to that of the BTZ black hole in Poincar\'{e} coordinates.   This is given by 
\begin{equation}
 ds^2 = \frac{1}{z^2} \left ( - ( 1-z^2) dt^2 + \frac{dz^2}{ 1- z^2} + d\phi^2 \right) .
\end{equation}
Now the solutions to the minimally coupled scalar in the BTZ black hole is known 
\cite{Birmingham:2001hc}. 
The two independent solutions are given by 
\begin{eqnarray}
 \Phi_{\omega, k }^{(1)} (z, t, \phi) &=&   e^{ - i \omega t + i k \phi} (1- z^2)^{-\frac{i {\omega}}{2}} z^{\Delta_-} 
 \\ \nonumber
& & \times F \left(\frac{1}{2} (\Delta_- - i({\omega} - k)),\frac{1}{2}(\Delta_- - i ({\omega} + k)),1-i {\omega}, 1- z^2\right),
\\ \nonumber
\Phi_{\omega, k }^{(2)} ( z, t, \phi) &=&  e^{ - i \omega t + i k \phi}(1- z^2)^{\frac{i {\omega}}{2}} z^{\Delta_-}  \\ \nonumber
& & 
\times 
F \left(\frac{1}{2} (\Delta_- + i ({\omega} + k)),\frac{1}{2} (\Delta_- + i({\omega} - k)),1+ i {\omega}, 1- z^2 \right).
\end{eqnarray}
Note that these two independent solutions reduce to  the ingoing and outgoing Fourier modes  at the horizon. 
We can now obtain the solution in the coordinates $(z, v, \phi)$ by performing the substitution given in (\ref{vtos}). 
This leads to the following independent solutions for the Vaidya metric  in the region $v>0$,
\begin{eqnarray} \label{btzsol-horizon}
  \Phi_{\omega, k }^{(1)} (z, v, \phi) &=& e^{ - i \omega v + i k \phi} (1+ z)^{ i \omega } z^{\Delta_-} 
   \\ \nonumber
& & \times F \left(\frac{1}{2} (\Delta_- - i({\omega} - k)),\frac{1}{2}(\Delta_- - i ({\omega} + k)),1-i {\omega}, 1- z^2\right),
\\ \nonumber
\Phi_{\omega, k }^{(2)} ( z, v, \phi) &=&  e^{ - i \omega v + i k \phi}(1- z)^{ i \omega } z^{\Delta_-}   \\ \nonumber
& & \times 
F \left(\frac{1}{2} (\Delta_- + i ({\omega} + k)),\frac{1}{2} (\Delta_- + i({\omega} - k)),1+ i {\omega}, 1- z^2 \right).
\end{eqnarray}
Now the above solutions admit an expansion around the horizon $z=1$, however we have seen in section \ref{sec-gen},  
to obtain the Green  function we need an expansion around the boundary $z=0$. 
This can be achieved by using the transformation properties of the hypergeometric functions. 
We have performed the required transformation in appendix A.  This results in the following 
independent solutions,
\begin{eqnarray}  \label{btzsol}
 \Phi_{\omega, k}^{BTZ (+)} (z, v, \phi) & =& e^{ - i \omega v + i k \phi} z^{\Delta_-} 
 ( 1+ z)^{i \omega}  \times \\ \nonumber
& & F \left(\frac{1}{2} (\Delta_- - i(\omega - k)),\frac{1}{2}(\Delta_- - i (\omega + k)),\Delta_-,  z^2\right),
\\ \nonumber
  \Phi_{\omega, k}^{BTZ (-)} (z, v, \phi) & =& e^{ - i \omega v + i k \phi} z^{\Delta_+} 
  ( 1+ z)^{i \omega}  \times \\ \nonumber
& &  F \left(\frac{1}{2} (\Delta_+ - i(\omega + k)),\frac{1}{2}(\Delta_+ - i (\omega - k)),\Delta_+, z^2\right).
\end{eqnarray}

It can be explicitly verified  that the
above solutions satisfy the equations of motion given in (\ref{ads3-4}) for $v>0$. 
The partial Fourier transform of this solution with respect to  $\omega$ is given by 
\begin{eqnarray} \label{ads3-8}
&& \Phi^{BTZ}(k,v,z)\\
= && z^{\Delta_-} \int_{-\infty}^{\infty} d\omega \, e^{- i \omega v} \, C_1(\omega) (1+ z)^{-i \omega} 
F \left(\frac{1}{2} (\Delta_- - i(\omega - k)),\frac{1}{2}(\Delta_- - i (\omega + k)),\Delta_-,  z^2\right) \nonumber\\
+ && z^{\Delta_+} \int_{-\infty}^{\infty} d\omega \, e^{- i \omega v} \, C_2(\omega) (1+ z)^{-i \omega} 
F \left(\frac{1}{2} (\Delta_+ - i(\omega + k)),\frac{1}{2}(\Delta_+ - i (\omega - k)),\Delta_+, z^2\right). \nonumber
\end{eqnarray}
The above solution admits an expansion around the boundary $z=0$.  Therefore we  have 
 written the solution  in the black hole region in the  required form given in (\ref{gen-0-1}).  Note that 
 in the case of the BTZ black hole, the Frobenius expansion around the boundary can be written in closed form.

\subsection{Matching at $v=0$ and Green function}

We follow the general procedure discussed in section  \ref{sec-gen} to construct the time dependent Green function. 
To do this we first obtain the moments of the function $C_2(\omega)$  by matching  the wave 
function $\Phi^{AdS}( v, k, z)$  in (\ref{ads3-5})  and $\Phi^{BTZ}( v, k , z) $ in  (\ref{btzsol})  at $v=0$. 
The moments of $C_2(\omega)$ are determined by comparing powers of $z$  in the terms proportional to 
$z^{\Delta_+}$. 
We will  demonstrate this procedure explicitly  and obtain moments up to the second order.  
We expand  the LHS of 
(\ref{ads3-5})   and the $z^{\Delta_+}$ coefficient  of (\ref{btzsol}) to 
quadratic order in $z$ to obtain the following equation
\begin{equation}
\begin{split}
& C \theta(-t_1) \left(\frac{|k|}{|t_1|}\right)^{\nu+\frac{1}{2}} \left(1 + \left(\frac{\nu}{2} + \frac{1}{4}\right) \left( \frac{2 z}{t} \right) + \frac{1}{2!}\left(\frac{\nu}{2} + \frac{1}{4}\right)\left(\frac{\nu}{2} + \frac{5}{4}\right)\left( \frac{2 z}{t} \right)^2 \dots \right) \times\\
& \left( J_{-\nu-\frac{1}{2}}(|k||t_1|) + z \frac{d J_{-\nu-\frac{1}{2}}\left( |k| \sqrt{t_1^2-2t_1z} \right)}{dz} \big|_{z=0}+ \frac{z^2}{2} \frac{d^2 J_{-\nu-\frac{1}{2}}\left( |k| \sqrt{t_1^2-2t_1z} \right)}{dz^2} \big|_{z=0} + \dots \right)\\
= & \int_{-\infty}^\infty d\omega C_2(\omega) (1-i \omega z + i \omega(i \omega+1) \frac{z^2}{2} + \dots) \times\\
& \times \big( 1 + \frac{\frac{1}{2} (1+\nu - i(\omega + k))\frac{1}{2} (1+\nu - i(\omega - k))}{1+\nu}z^2 + \dots \big)\\
= & \int_{-\infty}^\infty d\omega C_2(\omega) (1-i \omega z + \left(\frac{1+\nu}{4} - \frac{\omega^2}{2} - \frac{\omega^2-k^2}{4 (1+\nu)}\right) z^2 + \dots).
\end{split}
\end{equation}
Comparing the terms we can read out the moments to quadratic order.  These are given by 
\begin{eqnarray}
M_0 && = \int_{-\infty}^\infty d\omega \, C_2(\omega) = N J_{-\nu-\frac{1}{2}}(|k||t_1|),\\
M_1 && = \int_{-\infty}^\infty d\omega \, w C_2(\omega) = -i N k J_{-\nu-\frac{3}{2}}(|k||t_1|), \nonumber\\
M_2 && = \int_{-\infty}^\infty d\omega \, w^2 C_2(\omega) = N \Big[\left( \frac{(1+\nu)^2}{3+2\nu}+k^2 -\frac{2(1+\nu)(1+2\nu)}{t_1^2}\right) J_{-\nu-\frac{1}{2}}(|k||t_1|) \nonumber\\
&& \qquad \qquad \qquad \qquad \qquad \qquad - \frac{2 k (1+\nu)}{t_1} J_{-\nu+\frac{1}{2}}(|k||t_1|) \Big],\nonumber
\end{eqnarray}
here, $N \equiv C \theta(-t_1) \left(\frac{|k|}{|t_1|}\right)^{\nu+\frac{1}{2}}$.
As discussed in section  \ref{sec-gen}, see equation (\ref{greenp}),   the  retarded Green  function is given by 
\footnote{Since we are interested only in the time dependence we are ignoring  overall  proportionality constants
in the Green function. }
\begin{equation}
 G(v) = M_0 + i v M_1  - \frac{1}{2!} v^2 M_2 + \cdots  .
\end{equation}
Thus it is clear that the short time expansions of the Green function can be easily obtained. 

It is interesting to take the $k=0$ limit of the moments and construct the short time expansion of the Green  function. 
In this limit the Bessel function reduces to a rational function and  the first few moments are given by 
\begin{align} \label{kzero}
\begin{split}
& M_0 = \frac{C}{t_1^{2 \nu + 1}} \frac{2^{\nu+\frac{1}{2}}}{\Gamma(-\nu+\frac{1}{2})},\\
& M_1 = \frac{C}{t_1^{2 \nu + 1}} \frac{ 2^{\nu+\frac{1}{2}}}
{\Gamma(-\nu+\frac{1}{2})} \frac{-i \, 2(-\nu +\frac{1}{2})}{t_1},\\
& M_2 = \frac{C}{t_1^{2 \nu + 1}} \frac{2^{\nu+\frac{1}{2}}}
{\Gamma(-\nu+\frac{1}{2})} \frac{1+\nu}{(3+2\nu)} \left(1+\nu-\frac{2}{t^2_1}(1+2\nu)(3+2\nu) \right).
\end{split}
\end{align}
Note that it is clear that the Green function does not vanish identically for $k=0$ for arbitrary $t_2, v$, since these moments 
do not vanish.

\subsection{Recursive  numerical construction of the Green function}

It is easy to set up an algorithm in Mathematica to evaluate the moments recursively as discussed in 
section \ref{sec-gen}.  This algorithm is used to  evaluate $56$ moments of the function $C_2(\omega)$. 
From (\ref{greenp}) we can construct  the Green  function to $O(v^{56})$ \footnote{We stopped at this 
order in moments since we found that for times $v=5$ in horizon units  the Green  function 
evaluated converged to high degree of accuracy.}
To improve accuracy we then approximate the Green function using  the  $(28|28)$th   Pad\'{e}  approximant. 
The results for the Green  function are given in the 3 figures which we will discuss.

\begin{figure}
\centering
\includegraphics[scale=1]{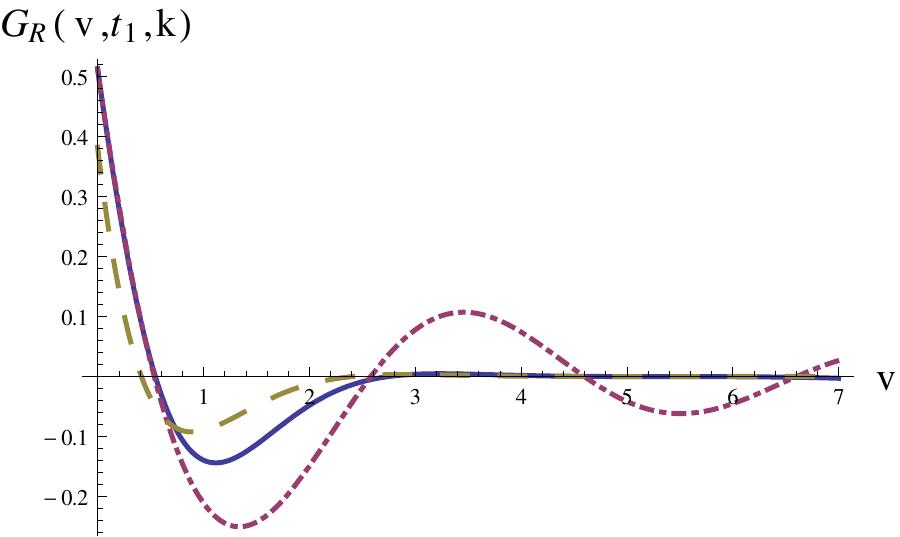}
\caption{The thermalizing (solid blue curve), vacuum (dot-dashed red curve) and thermal (dashed green curves) 
Green functions are plotted as a function of future time, $v$, for fixed values, $\nu = \frac{2}{3}, k = \frac{\pi}{2}$ and $t_1 =- 1.7$.}
\label{fig1}
\end{figure}

Figure \ref{fig1}, shows the thermalizing Green function (solid blue curve), the vacuum Green  function 
(dot-dashed red curve), and the thermal Green  function (green dashed curve), as a function of future time $v$. 
The thermalizing Green  function starts close to the vacuum Green  function, for small time, however it deviates away 
from the vacuum Green  function within one horizon time. At large time, the thermalizing Green  function approaches the 
thermal one. At $v=0$, the thermal Green function starts at a different value than the thermalizing and the vacuum Green functions. The thermal Green function is plotted from (\ref{thermal}) derived in the appendix \ref{app-thermal}. Figures \ref{fig1} and \ref{fig2}, are plotted for specific values, $\nu = \frac{2}{3}, k = \frac{\pi}{2}$ and $t_1 = -1.7$.

Figure \ref{fig2}  is the logarithmic plot of the absolute value of the thermalizing Green  function 
(solid blue curve), and the imaginary part of the lowest quasinormal mode (dot-dashed red curve), $e^{- \Delta_+ v}$. It is seen that for large time, the decay of the thermalizing Green function is given by this lowest quasinormal mode. Here, large time means time of the order of a few horizon radius, as can be seen from the plot. 
It is important to note that thermalization, i.e. decay of thermalizing Green function follows the lowest quasinormal mode, 
is achieved within a few ($ \sim O(1)$) horizon radius. The dips in the 
plot are the points where the thermalizing Green function crosses the time axis in figure \ref{fig1}, and indicates the oscillations 
of this Green function. Figures \ref{fig1} and \ref{fig2} reproduce those found in  \cite{Balasubramanian:2011ur} 
by the direct numerical integration of the 
differential equation (\ref{ads3-4}).

Lastly, in figure \ref{fig3}, the thermalizing (red dots) and the thermal (solid blue curve) Green function are plotted as function of $k$. 
The value of the thermalizing Green function close to $k =0$ is non zero, as expected from the earlier discussion in around (\ref{kzero}). 
Also, the 
values of the thermalizing and the thermal Green functions are close to each other near $k=0$. 
This figure is plotted for specific 
values, $\nu = \frac{2}{3}, t_1 = -2$ and $t_2=5$.

\begin{figure}
\centering
\includegraphics[scale=1]{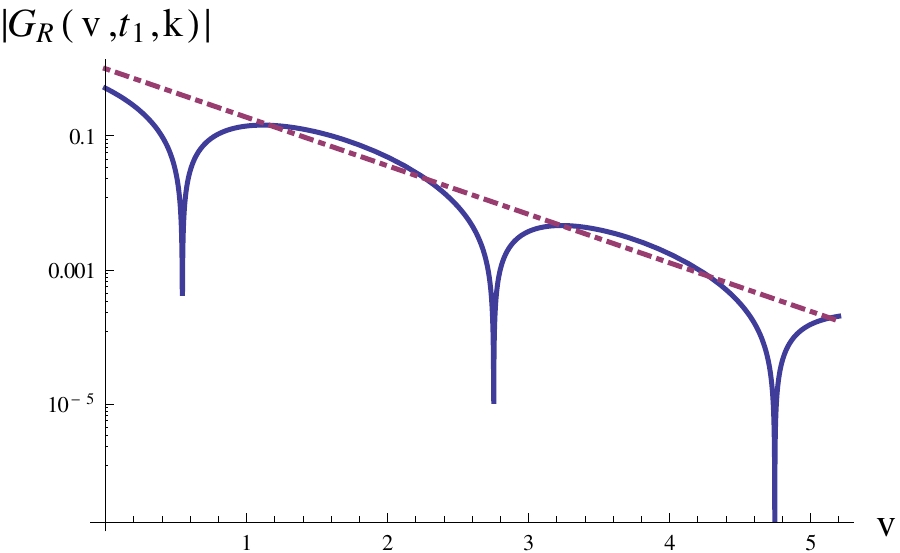}
\caption{The logarithmic plot of the absolute value of the thermalizing Green
function is given by the solid blue curve, for fixed values, $\nu = \frac{2}{3}, k = \frac{\pi}{2}$ and $t_1 = -1.7$. 
The dot-dashed red line gives the lowest quasinormal mode of the thermal Green function.}
\label{fig2}
\end{figure} 

\begin{figure}
\centering
\includegraphics[scale=1]{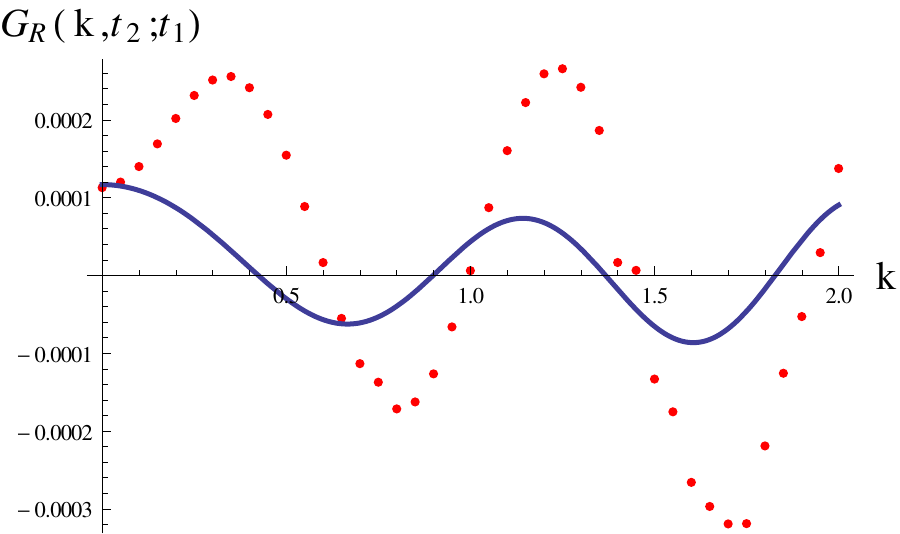}
\caption{The thermalizing and thermal Green functions as functions of $k$ are given by the 
red dots and solid blue curve, respectively, for fixed values, $\nu = \frac{2}{3}, t_1 = -2$ and $t_2=5$. 
From the plot it is seen that values of the thermalizing and thermal Green function are close to each other near $k=0$.}
\label{fig3}
\end{figure} 

\section{Long time behaviour of the Green function} \label{longt}

It is clear from the expression for the Green function given in (\ref{greenp}), that for obtaining the long time 
behaviour of the Green function we need to evaluate  moments $M_n$ for large values of $n$.  
Form the recursive relations for the moments in (\ref{recurm}) we see that this can be  done if we implement the matching 
of the wave functions to order $z^n$.  This in turn implies that we need the knowledge of the  less dominant  wave 
function in (\ref{gen-0-1})   $\Phi^{+}$  defined by 
\begin{equation}
\Phi^+( \omega,  k, z) =  \sum_{n=0}^\infty z^{\Delta_+}  A_n ( \omega, k, z) z^n,  
\end{equation} 
closer to the horizon. This is because the coefficients of $z^n$  for large values of $n$ will be determined by the 
singular behaviour at the horizon. 
The behaviour of this wave function near the horizon  can be determined by the general properties of 
solutions of wave equations in $AdS$ black holes.  
Using this information we will show that the long time behaviour of the 
Green function is determined by the first quasinormal mode. 
We will  first demonstrate this for the case of the minimally coupled scalar in $AdS_3$ Vaidya. 
This is easy to do explicitly since,  the wave function $\Phi^{BTZ (+)}$ is known in closed form. 
We will then  show in general that the long time behaviour of the Green function is determined by the 
first quasinormal mode of the wave functions in the corresponding black hole background.

\subsection{Green function in   $AdS_3$ Vaidya } \label{lt3}

From the preceding discussion we need to examine the behaviour of the function $\Phi^+(\omega, k , z)$ close to the horizon. 
For the case of the minimally coupled scalar in the  BTZ black hole this function is known in closed form (\ref{ads3-8}) and is given by 
\begin{equation}
 \Phi^{BTZ(+)}(\omega, k , z)  = z^{\Delta_+} ( 1+ z)^{-i\omega} {}_2F_1 \left( 
 \frac{1}{2} ( \Delta_+ - i ( \omega + k) ) , \frac{1}{2} ( \Delta_+ - i ( \omega - k) ), \Delta_+ , z^2 
 \right).
\end{equation}
The near horizon limit of this wave function can be easily determined from the properties of the 
hypergeometric function given in (\ref{conformula}). 
We see that the singular behaviour near $z=1$ is given by 
\begin{equation} \label{nearsingb}
\Phi^{BTZ(+)}(\omega, k , z)  \sim  ( 1- z)^{i \omega} 
\frac{\Gamma(  \Delta_+ ) \Gamma( -i\omega) }{ \Gamma \frac{1}{2}( \Delta_+  - i ( \omega - k) ) 
\Gamma( \frac{1}{2} (  \Delta_+  - i ( \omega + k) ) }.
\end{equation}
This equation will serve as the starting point of obtaining the long term
behaviour of the Green function. It is important  to note that 
the mode $\Phi^+$ vanishes near the horizon when the frequency is given by 
\begin{equation} \label{qnm}
  \omega_n^\pm  = - i \Delta_+ - 2ni  \pm  k, \qquad n = 0 , 1, 2, \cdots .
 \end{equation}
 These are the quasinormal modes of the minimally coupled scalar in the BTZ black hole \cite{Birmingham:2001pj}. 
 Also note that these modes are located in the lower half $\omega$-plane. 
We now impose the matching condition at $v=0$.  This will lead to the following equation
\begin{eqnarray}\label{matchh}
& &C \frac{|k|^{\nu + \frac{1}{2}} }{ (  t_1^2 + 2  |t_1|  z ) ^{\frac{( 2\nu + 1) }{4}} }
J_{- \nu - \frac{1}{2} } ( |k| \sqrt { t_1^2 + 2  |t_1|  z } )   \\ \nonumber
&\sim  & \int_{-\infty}^{\infty} d\omega   C_2 ( \omega) 
  ( 1-z)  ^{-i\omega} 
\frac{\Gamma(  \Delta_+  ) \Gamma( -i\omega) }{ \Gamma \frac{1}{2}( \Delta_+  - i ( \omega - k) ) 
\Gamma( \frac{1}{2} ( \Delta_+  - i ( \omega + k) ) }.
\end{eqnarray}
We expect this approximation near the horizon to estimate the behaviour of the moments suitable to 
obtain long time behaviour of the Green function.  
We can now solve for $C_2(\omega)$ by substituting 
\begin{equation}
 y = \ln( 1-z) , 
\end{equation}
and multiplying both sides of the equation in (\ref{matchh}) by $e^{-i \omega' y}$ and formally integrating 
over $y$ from $-\infty$ to $\infty$ \footnote{Though the range of $y$ is restricted from $-\infty$ to $0$, 
we are extending this range formally to obtain $C_2$. }. 
This leads to the following expression for $C_2(\omega)$ 
\begin{eqnarray}
C_2 (\omega)  \sim  \tilde I (\omega, k )  
\frac{ \Gamma \frac{1}{2}( \Delta_+  - i ( \omega - k) ) 
\Gamma( \frac{1}{2} ( \Delta_+ - i ( \omega + k)) }{\Gamma(  \Delta_+   ) \Gamma( -i\omega) },
  \end{eqnarray}
 where 
  \begin{eqnarray}
  \tilde I(\omega, k ) =  \frac{C}{2\pi} \int_{-\infty}^\infty dy \frac{ |k|^{\nu + \frac{1}{2}}}
  {  (t_1  ^2 + 2 |t_1| z)  ^{ \frac{  2\nu + 1}{ 4}  } }
  J_{-\nu - 1/2} ( |k| \sqrt{  t^2 + 2 | t_1| z} )  e^{ - i \omega y} . \nonumber \\
  \end{eqnarray}
 Finally the Green function is given by 
  \begin{eqnarray} \label{contint}
 & & G_R( k, v,  t_1 ) =  \int_{-\infty}^\infty d\omega e^{-i\omega v} C_2(\omega) = \\ \nonumber
 & & \int_{-\infty}^\infty d\omega  e^{-i\omega v} 
  \frac{ \Gamma \frac{1}{2}( \Delta_+  - i ( \omega - k) ) 
\Gamma( \frac{1}{2} ( \Delta_+ - i ( \omega + k)) }{\Gamma(  \Delta_+   ) \Gamma( -i\omega) }
\tilde I(\omega, k ) .
\end{eqnarray}
Note that the factor multiplying $\tilde I(\omega, k ) $ in  (\ref{contint})   is a 
function of $\omega$  such that for  $\omega \rightarrow - i  \infty $  
it behaves as 
\begin{eqnarray} \label{bounlim}
 \lim_{\omega \rightarrow - i\infty}  H(\omega, k )  && =  
 \lim_{\omega \rightarrow - i\infty} \frac{ \Gamma \frac{1}{2}( \Delta_+  - i ( \omega - k) ) 
\Gamma( \frac{1}{2} ( \Delta_+ - i ( \omega + k)) }{\Gamma(  \Delta_+   ) \Gamma( -i\omega) }, \\ \nonumber
& & \sim 
e^{ + i\omega  \ln 2} ( \omega) ^{ \nu + \frac{1}{2} } .
\end{eqnarray}
$\tilde I(\omega, k ) $ is essentially a Fourier transform of the Bessel function. We   assume  that this is
analytic in the lower half $\omega$-plane and grows {\it at the most}  exponentially in $\omega$ as 
$\omega \rightarrow - i\infty$,  given by 
\begin{equation}
 \lim_{\omega \rightarrow - i\infty} \tilde I (\omega, k)  < e^{ i \omega M} .
\end{equation}
Here $M>0$ is a fixed constant. 
With these assumptions and the behaviour in (\ref{bounlim}) we see that integrand in 
(\ref{contint}) goes to zero in the limit $\omega \rightarrow - i\infty$ for sufficiently large but fixed  $v>0$. 
This is because of the exponentially dying term 
 $e^{-i\omega v} $ in the integrand of (\ref{contint}). 
 Therefore the integral can be performed by completing the contour in the lower half $\omega$-plane. 
 Then the integral localizes to a sum over the poles of the Gamma functions.  
 The poles of the Gamma functions are at the quasinormal modes given by (\ref{qnm}).  
The result of the integral then reduces to 
\begin{eqnarray}\label{finallong}
G ( k, v, t_1)   \sim \sum_{n=0, \alpha = \pm }^{\infty} e^{ - i\omega_n^\alpha v } \tilde I ( \omega_n^\alpha , k )
( - 2\pi i \, {\rm Residue}_{\,  \omega = \omega_n^{\alpha}}\,   H(\omega, k ) )  .
 \end{eqnarray}
It is now clear from (\ref{finallong}) 
that the long time behaviour of the Green function is determined by the lowest quasinormal mode as we have seen 
in our explicit numerical evaluation of the Green function.  The decay of the Green function is 
controlled by the imaginary part of the lowest quasinormal mode and the period of oscillations is 
determined by  the real part of the lowest quasinormal mode. 
Note also from (\ref{finallong}) we see that since the expression  involves  a sum over all the quasinormal modes, the rough time 
scale over which the lowest quasinormal modes takes over  is of the order of a few horizon times. 
This is also clearly seen in the numerical evaluation of the Green function. 
It is important to note that the starting point of the analysis was the behaviour of 
$\Phi^{BTZ(+)}$  near the horizon given in (\ref{nearsingb}), which vanished at frequencies 
determined by the quasinormal mode spectrum in the lower half $\omega$-plane.

\subsection{Green function in $AdS_{d+1}$ Vaidya}

Using the intuition gained by the explicit solutions of the minimally coupled scalar in the BTZ black hole we now
generalize the discussion. 
We show that the long time behaviour of the retarded Green function in arbitrary $AdS_{d+1}$ Vaidya background is 
determined by the lowest quasinormal mode of the corresponding bulk field. 

Consider the differential equation given in (\ref{mincoup}) for the minimally coupled scalar 
in the $AdS_{d+1}$ Vaidya for $v>0$. Substituting 
\begin{equation}
  \phi ( v, k, z) = \Phi_\omega( k, z) e^{- i \omega v   },
\end{equation}
we obtain
\begin{equation}
 ( 1- z^d) \partial_z^2 \Phi_\omega  - i\omega \left(  \frac{ \Phi_\omega}{z}  - 2 \partial_z \Phi_\omega \right)
-  \left( d z^{d-1}  +  \frac{1}{z} ( 1- z^d)  \right) \partial_z \Phi_\omega - 
 \left( \frac{m^2}{z^2} + k^2 \right) \Phi_\omega = 0 .
\end{equation}
This equation has two regular singular points,
$z=0$ and $z=1$, corresponding to the boundary and the horizon of the black hole. 
One can set up a Frobenius solution around either $z=0$ or $z=1$. 
Let the two independent solutions around $z=0$ be $\Phi_\omega^+$ and $\Phi_\omega^-$.  From the indicial equation 
for the expansion at $z=0$ we know that these solutions  behave as follows \footnote{We have assumed that the roots of the 
indicial equation do not differ by an integer.  The discussion can be easily generalized in case they do.}  
\begin{equation}
 \lim_{z\rightarrow 0} \Phi_\omega^+ (z, k ) \sim z^{\Delta_+},  \qquad
 \lim_{z\rightarrow 0} \Phi_\omega^-  (z, k) \sim z^{\Delta_-} .
\end{equation}
Therefore $\Phi_\omega^-$ is the dominant  singular mode at the boundary. 
Similarly one can set up an expansion around $z=1$, let the two independent solutions around $z=1$ be $\Phi_{\omega}^{in}$ and 
$\Phi_{\omega}^{out}$.  It is easy to see from the indicial equation around $z=1$, that their behaviour 
near $z=1$ is given by 
\begin{equation}\label{sinnear}
 \lim_{z\rightarrow 1}\,  \Phi_{\omega}^{out} (z, k ) \sim e^{i \frac{2 \omega}{d}  \ln( 1-z) } , \qquad
 \lim_{z\rightarrow 1}\,  \Phi_{\omega}^{in}(z, k) \sim   (1-z)^{0}. 
\end{equation}
The reason we have labeled these solutions as ${in}$ and $out$ is because they correspond to the ingoing and 
outgoing solutions when these wave functions are transformed to the $t, z$ coordinates. 
To see this note that the coordinate transformation near $z=1$ can be obtained by integrating 
(\ref{difftrans}). 
This is given by 
\begin{equation}
 v \sim  t + \frac{1}{d} \ln (1-z) .
\end{equation}
Substituting this coordinate transformation we obtain the corresponding wave functions in the $t, z$ coordinates
\begin{eqnarray}
 \phi^{out}_\omega ( v, z) = e^{ - i\omega v } \Phi_{\omega}^{in} (z, k) \sim  e^{-i\omega t} ( 1-z) ^{\frac{i\omega}{d} }, 
 \\ \nonumber
 \phi^{in}_\omega(v, z) = e^{-i\omega v} \Phi_{\omega}^{out} (z, k) \sim e^{- i \omega t} ( 1-z) ^{- \frac{i \omega}{d}} .
\end{eqnarray}

Now we have 2 sets of linearly independent solutions $\{ \Phi_\omega^+, \Phi_\omega^- \} $ and 
$\{ \Phi_\omega^{in}, \Phi_{\omega}^{out} \}$.  Therefore by the uniqueness theorem of second order 
ordinary linear differential equations we should be able to express one  set in terms of the other as linear combinations.  
Let us write  
\begin{eqnarray}\label{matrix}
 \Phi_\omega^{out} (k, z) &=& M_{11}(\omega, k)  \Phi_\omega^{+}(k, z)  + M_{12}(\omega, k ) \Phi_{\omega}^{-}(k, z) ,
 \\ \nonumber
 \Phi_\omega^{in} (k, z) &=&  M_{21}(\omega, k)  \Phi_\omega^{+}(k, z)  + M_{22}(\omega, k ) \Phi_{\omega}^{-}(k, z) .
\end{eqnarray}
Let us now use the definition of quasinormal modes to obtain some information of the  coefficients $M_{12}$ and $M_{22}$. 
We have seen that $ \Phi_\omega^{in}, \Phi_\omega^{out}$ correspond to the ingoing and outgoing modes in the $t, z$ coordinates. 
Now by definition quasinormal modes are those values of frequencies for which  these modes obey Dirichlet boundary conditions 
at the horizon. This implies that the more dominant mode proportional to $\Phi_{\omega}^{-}$ should vanish 
at these frequencies. 
Thus we have  the equation
\begin{equation}
 M_{12} (\omega_n^{out}, k) = 0 , \qquad M_{22}( \omega_n^{in}, k ) = 0 ,
\end{equation}
where $\omega_n^{out}$ are the quasinormal frequencies  which lie in the upper half $\omega$-plane for the outgoing modes
and  $\omega_n^{in}$ are the quasinormal frequencies which lie in the lower half $\omega$-plane for the ingoing modes 
\cite{Birmingham:2001pj,Kovtun:2005ev}. 
Let us invert the equation (\ref{matrix}), we obtain
\begin{equation}
 \left( \begin{array}{c}
   \Phi_\omega^{+}(k, z) \\  \Phi_{\omega}^{-}(k, z) 
  \end{array}\right)   = \frac{1}{{\rm det} M } 
\left(  \begin{array}{cc}
  M_{22}   & - M_{12}  \\ - M_{21} & M_{11} 
 \end{array}\right)
 \left(
 \begin{array}{c}
   \Phi_\omega^{out} (k, z) \\  \Phi_\omega^{in} (k, z)
 \end{array}\right).
\end{equation}
Note that the inverse exists that is ${\rm det} M \neq 0$. This is because  $\Phi_\omega^+$ and $\Phi_\omega^-$ can be written 
as linear combinations of $\Phi_\omega^{in}$ and $\Phi_\omega^{out}$. 
We can now easily read out the near horizon behaviour of the solution $\Phi_{\omega}^{+}$ from 
the above equation.  The singular behaviour of $\Phi_{\omega}^{+}$  is given by 
\begin{eqnarray}
&& \Phi_\omega^{+}(k, z)  = \frac{1}{{\rm det} M} \left( M_{22}(\omega, k )  \Phi_\omega^{out}(k, z)     -
 M_{12} (\omega, k ) \Phi_{\omega}^{in}(k, z) \right) ,
\\ \nonumber
 && \Phi_\omega^{+}(k, z)|_{z\rightarrow 1} \sim  \frac{1}{{ \rm det M} } M_{22}(\omega, k )  ( 1-z)^{i\frac{2\omega}{d} }  .
\end{eqnarray}
Here we have used the near horizon behaviour given in (\ref{sinnear}). 
Now  $M_{22}(\omega, k )$ vanishes at $\omega = \omega_n^{in}$, with zeros in the lower 
half plane.  This is the general form of the equation (\ref{nearsingb}) seen explicitly for the case of the BTZ black hole for
black holes in $AdS_{d+1}$. 
Note that we arrived at this result from the general definition of quasi-normal modes. 
From this point  onwards we can follow the rest of the argument in section \ref{lt3} to arrive at the conclusion that 
the long time behaviour is determined by the leading quasinormal mode. 
In general the gaps between quasinormal modes are of the order of horizon scales. Therefore 
we expect the leading behaviour to set in order of a few horizon times.
Though here we have used the minimally coupled scalar to demonstrate our argument for simplicity, the analysis 
can be carried out for other fields.  The steps involved will result in similar equations. 
We will see this explicitly for the vector fluctuations of the metric in the subsequent section.

This concludes our general argument of why the long time behaviour is determined by the leading 
quasinormal mode. Thus any retarded correlator which is used to probe the onset of thermalization 
caused due to injection of  energy at an instance of time in the field theory will decay with the time 
scale set by the first quasinormal mode. In \cite{Horowitz:1999jd} it was conjectured that 
thermalization time scales in the field theory are determined by quasinormal modes. 
We have shown that the time dependent 
Green functions considered in this paper provides an explicit realization of this statement. 
From our general argument we expect this to be true for correlators whose lowest quasinormal modes
are determined by hydrodynamics if the probing momentum scales are  sufficiently small. 
In the next section we will verify this expectation for stress tensor correlators in ${\cal N}=4$ Yang-Mills 
whose lowest quasinormal mode is sensitive to the shear viscosity.

\section{Thermalization in $AdS_5$ Vaidya}

In this section we study the thermalization of the retarded two point functions in the thin shell 
$AdS_5$ Vaidya geometry. 
We first consider  the equation satisfied by the spin 2 metric fluctuation $h_{x^1 x^2}$.
This fluctuation, perpendicular to the momentum $k$ which is along the $x^3$ direction, forms the shear  spin 2 mode.  
Evaluating the retarded Green function holographically for this mode 
provides information of the retarded  two point function  of the stress tensor 
$\langle T_{x^1 x^2} T_{x^1 x^2} \rangle$,  for  the 
strongly coupled ${\cal N}=4$ Yang-Mills \footnote{Metric perturbations for spherical shell collapse was 
studied in \cite{Stricker:2013lma}.   Non-equilibrium Green function  corresponding 
to the shear correlator was studied earlier by 
a complementary approach  by examining backgrounds perturbed by hydrodynamic modes 
\cite{Banerjee:2012uq,Mukhopadhyay:2012hv}}. 
The lowest quasinormal mode of shear fluctuations of the metric has been studied earlier in 
\cite{Nunez:2003eq} numerically. 
We see our results are consistent with this earlier calculation. 
The more interesting metric fluctuation to consider is $h_{x^1 x^3}$. It is known that 
this mode admits a hydrodynamic quasinormal mode in the $AdS_5$ black hole
for  small momentum \cite{Policastro:2002se}.  This mode appears as a pole in the thermal  two 
point function  of the stress tensor $\langle T_{x^1 x^3} T_{x^1 x^3} \rangle$.  
We will  implement the  recursion method  numerically and  evaluate 
the  thermalizing Green function for various values of small momentum $k$. 
From the long time behaviour of the Green function we show that this Green function 
relaxes by the hydrodynamic quasinormal mode. 
This enables  us to read out  the universal shear viscosity to entropy density ratio from 
a time dependent process.

The thin shell $AdS$-Vaidya metric in $5$ dimensions is given by,
\begin{align}
\label{ads5-3} ds^2 &= \frac{1}{z^2} \left[-h(v,z)dv^2-2dzdv + d\vec{x}^2 \right],\\
& h(v,z) = 1-\theta(v) z^4,
\end{align}
here, $\vec{x}$ is a $3$ vector, with components, $x^i$, where $i=1,2,3$. 
The $x^i$'s parametrize the spatial directions in the dual field theory. 
The transformation which reduces the $v>0$ part of the above metric to the $AdS_5$ black hole metric in Poincar\'{e} co-ordinates is given by 
\begin{align}
\label{ads5-2} dv &= dt - \frac{dz}{1-z^4}.
\end{align}
In these coordinates we obtain the planar  black hole metric in $AdS_5$, for $v>0$,
\begin{align} \label{ads5-1}
ds^2 = \frac{1}{z^2} \left[-(1-z^4)dt^2+\frac{dz^2}{1-z^4} + d\vec{x}^2 \right].
\end{align} 
Here again the radius of $AdS$ as well as the black hole has been set to unity. 
The temperature of the black hole is then given by 
\begin{equation}
T= \frac{1}{\pi}.
\end{equation}

To study two point functions of the stress energy tensor of the boundary field theory
we   consider  small perturbations  to the background metric, 
$g^{(0)}_{\mu \nu}$, of (\ref{ads5-3}), $g_{\mu \nu} = g^{(0)}_{\mu \nu} + h_{\mu \nu}$. 
The differential equations satisfied by the metric perturbations are obtained by  linearizing
the Einstein's equations which is given by 
\begin{equation}
\mathcal{R}_{\mu \nu}  = - 4 \, g_{\mu \nu}.
\end{equation}
Here $\mathcal{R}_{\mu \nu}$ is the Ricci tensor and the value of cosmological constant, $\Lambda$, 
has been set to $\Lambda = -6$, in units of AdS radius. 
The linearized Einstein's equations are given by 
\begin{equation}\label{o1eom}
 \mathcal{R}_{\mu \nu}^{(1)}  = - 4 \, h_{\mu \nu},
\end{equation}
where $ \mathcal{R}_{\mu \nu}^{(1)} $ is the linearized Ricci curvature. 

\subsection{Spin 2  metric perturbations}

\begin{figure}
\centering
\begin{subfigure}{.5\textwidth}
  \centering
  \includegraphics[scale=0.8]{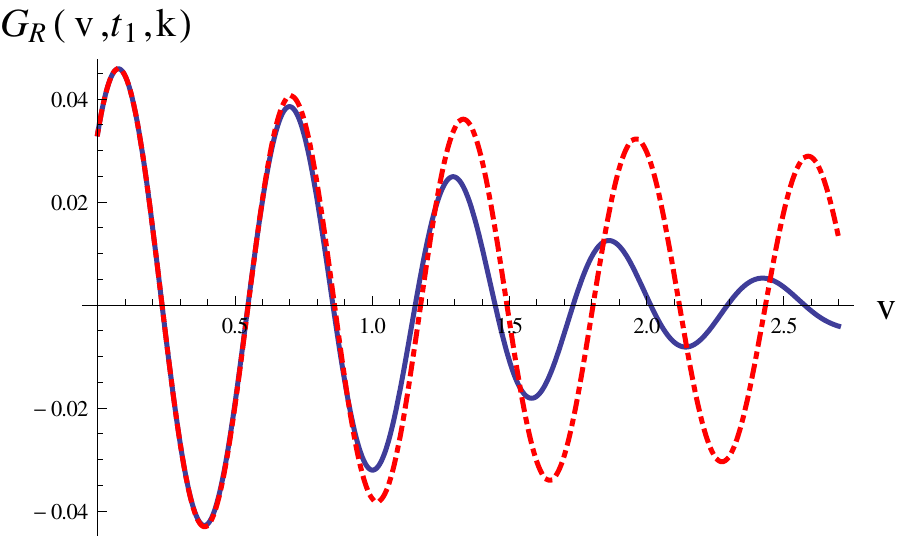}
  \caption{•}
  \label{fig-scalar-a}
\end{subfigure}%
\begin{subfigure}{.5\textwidth}
  \centering
  \includegraphics[scale=0.8]{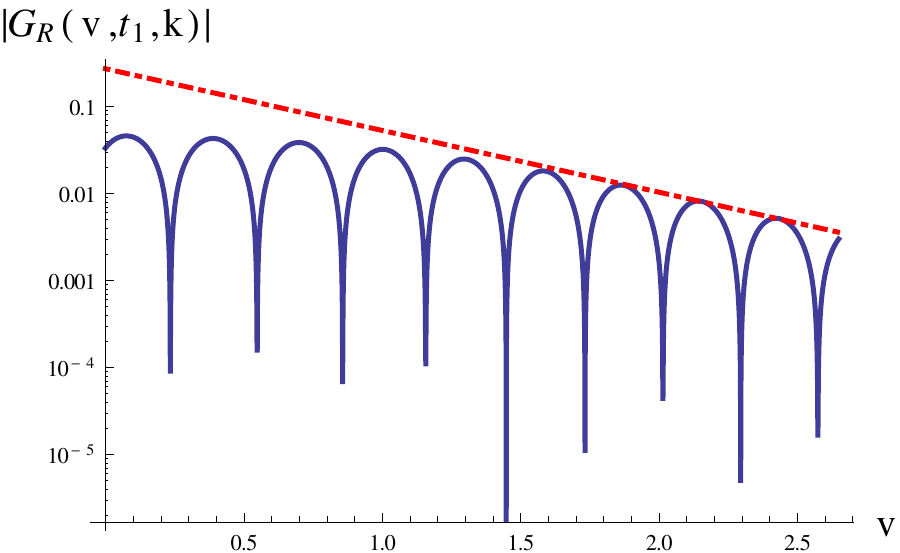}
  \caption{•}
  \label{fig-scalar-b}
\end{subfigure}
\caption{Figure (a) shows the thermalizing (solid blue curve), and vacuum (dot-dashed red curve) Green functions for the 
shear mode. Figure (b) shows the logarithmic plot of the absolute value of the thermalizing Green function. 
Plots are for values  $t_1 =- 15$ and $k=10$.}
\label{fig-scalar}
\end{figure}

First, we consider the case when only the scalar mode of metric perturbation, $h_{x_1 x_2}$, is turned on. 
We substitute 
 the ansatz $h_{x^1 x^2}(t,x^3,z)=e^{-i \omega t+i k x^3} \tilde{h}_{x^1 x^2}(z)$ in (\ref{o1eom}). Here, the momentum is chosen to be along the $x^3$ direction. 
 After the  redefinition   $\Phi(z) = z^2 \tilde{h}_{x^1 x^2}(z)$
 we obtain the  following equation from  (\ref{o1eom}),
\begin{align}
\label{ads5scalar1} v<0: & \quad z^2 \Phi'' - z(3-2 i \omega z) \Phi' - (k^2 z^2 + 3 i \omega z) \Phi = 0,\\
\label{ads5scalar2} v>0: & \quad z^2 (1-z^4) \Phi'' - z(3+z^4-2 i \omega z) \Phi' - (k^2 z^2 + 3 i \omega z) \Phi = 0,
\end{align}
where the prime means derivative with respect to $z$. Note that as expected these 
are the equations of the minimally coupled massless scalar in the $AdS_5$ background for 
$v<0$ and the $AdS_5$ black hole background for $v>0$. 
 
\paragraph{Solution for $v<0$}
\ \ \\

In appendix \ref{appendix_ads5}
 we have obtained the solution which satisfies the boundary condition (\ref{gen-0-0})
for a minimally coupled massive scalar in $AdS_5$. 
On  setting the parameter $m^2 =3$  or $\nu = 2$ in (\ref{scalarads}), 
the equation reduces to  (\ref{ads5scalar1}). Therefore we can read out the 
solution  in mixed Fourier space which satisfies the required boundary 
condition from (\ref{ads5bessel-sol}).     This is given by 
\begin{eqnarray}
&& \Phi^{AdS}(v,k,z;t_1)=G_R^{AdS}(v,k,z;t_1) =  \theta(v-t_1) 2^{-\frac{3}{2}}\sqrt{\pi}\times \\
&& \qquad \left(\frac{k}{\sqrt{(v-t_1)^2+2(v-t_1) z}}\right)^{\frac{5}{2}} z^4 J_{-\frac{5}{2}} \left(k \sqrt{(v-t_1)^2+2(v-t_1) z} \right) . \nonumber
\end{eqnarray}

\paragraph{Solution for $v>0$}
\ \ \\

Unlike the case of the BTZ black hole, closed form solutions to minimally coupled scalar 
in the $AdS_5$ black holes are not available. 
However the recursive method we developed in section \ref{sec-gen}   just relies on 
information of the Frobenius expansion of the solution near the boundary. 
The roots of the indicial equation of (\ref{ads5scalar2}) are given by $\Delta_\pm = 4,0$. 
As discussed earlier to obtain  the retarded 
Green function  we need to obtain  the Frobenius solution with the 
less dominant mode, that is with $\Delta_+ = 4$. 
The recursion relations for  the coefficients $a_n^j$s defined in (\ref{defan}), necessary for the construction of the 
solution in the region $v>0$, are given by  the following equations with  $a_0^0=1$,
\begin{eqnarray}\label{scalaran1}
&& a_1^0=0, \quad a_1^1=-ia_0^0, \\
&& a_2^0 = \frac{k^2 a_0^0}{12}, \quad a_2^1=0, \quad a_2^2 = \frac{-i 7 a_1^1}{12},\nonumber \\
&& a_3^0 = 0, \quad a_3^1=\frac{k^2a_1^1-i9a_2^0}{21}, \quad  a_3^2=0, \quad a_3^3 = \frac{-i9a_2^2}{21}, \nonumber
\end{eqnarray}
and, for $n \geq 4$,
\begin{eqnarray}\label{scalaran2}
&& a_n^0 =  \frac{k^2 a_{n-2}^0+n^2 a_{n-4}^0}{n(n+4)}, \\
&& a_n^j =  \frac{-i(2n+3) a_{n-1}^{j-1} +k^2 a_{n-2}^j+n^2  a_{n-4}^j}{n(n+4)}; \quad j=1,2, \dots, n-4,\nonumber\\
&& a_n^j = \frac{-i(2n+3) a_{n-1}^{j-1} +k^2 a_{n-2}^j}{n(n+4)}; \quad j=n-3,n-2, \nonumber\\
&& a_n^j = \frac{-i(2n+3) a_{n-1}^{j-1}}{n(n+4)}; \quad j=n-1,n.\nonumber
\end{eqnarray}
 
We can now extract the moments of the Green function by substituting  the values of $a_n^j$ determined from 
(\ref{scalaran1}),(\ref{scalaran2}) into (\ref{recurm}). 
The values of the coefficients $\tilde{J}_n$ in (\ref{recurm}) can be obtained from the recursion relation  (\ref{recurjn}) with $\nu=2$. 
Finally the retarded  Green function is constructed using (\ref{greenp}).  This procedure is clearly 
algorithmic and can be easily implemented numerically. 
Using Mathematica we implemented this procedure of obtaining the Green function. 
We have evaluated  up to 200 moments $C_2(\omega)$ and the Green function is constructed to $O(v^{200})$ in future time, $v$.
We then approximated the  Green function  by the $(100|100)$ Pad\'{e} Approximant for better accuracy.
For the values of $k$ studied, 200 moments were sufficient to obtain  convergent results. 

Figure \ref{fig-scalar-a} shows the thermalizing Green function (solid blue curve) for $t_1=-15$ 
and a particular value of $k=10$. As expected, the thermalizing Green function starts close to the vacuum 
Green function (red dot-dashed curve) at small time, and deviates from it within one horizon time. 
From the logarithmic plot, figure \ref{fig-scalar-b}  the slope as shown by the dot-dashed red line of the 
thermalizing Green function (solid blue curve) is measured at large time, i.e. at time of the order of a few horizon radius. 
This slope gives the negative imaginary part of the lowest quasinormal mode ($-$Im $\omega$), as shown in 
section \ref{longt}. From the gaps between consecutive zeroes of the thermalizing Green function, 
real part of $\omega$ can be calculated using, $ \text{Re} \, \omega = \pi/ \text{gap}$. 
The values of real and imaginary part of $\omega$, for several values of $k$, are listed in table \ref{table}. 
These values agree with figures 5,6 of \citep{Nunez:2003eq}  where the numerical values of the 
real and imaginary part of the quasinormal modes for the minimally coupled scalar with $\Delta =4$ in $AdS_5$ was 
obtained for various values of $k$. 
To compare our results to that of figures 5,6 of \citep{Nunez:2003eq}, note that 
we need to perform  the following scalings $ k^{ours} =  2 q^{theirs},  \omega^{ours} = 2 \omega^{theirs}$. 
As a simple check note that $k^{ours} =10$ corresponds to $q^{theirs} =5 $, looking at their figure 6 we note that ${\rm Im} \, \omega^{theirs} = .8$
which corresponds to ${\rm Im}\,  \omega^{ours} = 1.6$ which agrees with our result in table \ref{table}. 
Similarly note that for small values of $q^{theirs}$, the value of ${\rm Re}\,  \omega^{theirs}$ is above the 
$45^{\circ}$ line. This is also the case from our results in table \ref{table}. 

\begin{table}[h]
\begin{center}
\begin{tabular}{|l|r|r|}
\hline
$k$ & Re $\omega$ & $-$Im $\omega$ \\
\hline
7.0 & 8.34 & 1.90 \\
7.5 & 8.84 & 1.77 \\
8.0 & 9.29 & 1.76 \\
8.5 & 9.80 & 1.78 \\
9.0 & 10.28 & 1.76 \\
9.5 & 10.72 & 1.63 \\ 
10.0 & 11.21 & 1.63 \\
10.5 & 11.70 & 1.52 \\
11.0 & 12.22 & 1.55 \\
\hline
\end{tabular}
\end{center}
\caption{Values of the real and imaginary part of the lowest quasinormal modes, $-$Im $\omega$, 
calculated from the large time behaviour of the thermalizing Green function, for several values of momentum, 
$k$, and fixed value of $t_1=-15$.}
\label{table}
\end{table}

\subsection{Vector metric perturbations  and shear viscosity}

It is known that  vector metric  perturbations $h_{x^1 x^3}, h_{t x^1}$ in the $AdS_5$ black hole 
with momentum is along the $x^3$ direction admit a hydrodynamic mode at small momentum in addition 
to the usual gapped quasinormal frequencies \cite{Policastro:2002se}.  The hydrodynamic quasinormal  mode corresponds 
to the hydrodynamic pole in the thermal correlator $\langle T_{t x^1} T_{t x^1} \rangle$. 
This was used to read out the ratio of shear viscosity to entropy density in \cite{Policastro:2002se}. 
Using the methods developed in this paper we can evaluate the time dependent  thermalizing 
retarded  two point function $\langle T_{t x^1} T_{t x^1} \rangle$ in the $AdS_5$ thin shell Vaidya background. 
From  the general analysis of section \ref{longt} we expect that the time dependent 
Green function in the $AdS_5$ thin shell Vaidya background  should relax to 
equilibrium by the hydrodynamic quasi-normal mode. 
Therefore from the decay it should be possible to read out the ratio of shear viscosity to 
entropy density from a dynamical Green function. 
In this section we perform this analysis using the methods developed in this paper 
and obtain the universal ratio of shear viscosity to entropy density\footnote{As far as the authors are aware
this is the first instance where the universal  ratio of shear viscosity to entropy density is obtained from 
a time dependent process. }.

We first turn on the following metric fluctuations with momentum along the $x^3$ direction,
\begin{equation}
h_{v x^1}(t,x^3,z)=e^{-i \omega t+i k x^3}\tilde{h}_{v x^1}(z), \qquad 
h_{x^1 x^3}(t,x^3,z)=e^{-i \omega t+i k x^3} \tilde{h}_{x^1 x^3}(z).
\end{equation}
We can obtain the linearized equations of motion from 
(\ref{o1eom}). 
Redefining the fields as, 
\begin{equation}
H_v(z)=z^2 \tilde{h}_{v x^1}(z), \qquad H_3(z)=z^2 \tilde{h}_{x^1 x^3}(z),
\end{equation}
we obtain the following  coupled equations in the $AdS_5$ background before  the collapse 
of the thin shell,  $v<0$,
\begin{align}
& H''_v - \frac{3}{z} H'_v + i k H'_3=0, \\
& H''_v+ \left(-\frac{3}{z}+i\omega \right) H'_v-k^2 H_v -k \omega H_3=0, \nonumber\\
& H''_3 + \left(-\frac{3}{z} + 2 i \omega\right) H'_3-\frac{3i\omega}{z} H_3+i k H'_v -\frac{3ik}{z} H_v=0. \nonumber
\end{align}
After the formation of the $AdS_5$ black hole for $v>0$, the equations are given by  
 \begin{align}
& H''_v - \frac{3}{z} H'_v + i k H'_3=0,\\
& H''_v+ \frac{1}{f}\left(-\frac{3-3z^4}{z}+i\omega \right) H'_v-\frac{k^2}{f} H_v - \frac{k \omega}{f} H_3=0, \nonumber\\
& H''_3 + \frac{1}{f} \left(-\frac{3+z^4}{z} + 2 i \omega\right) H'_3-\frac{3i\omega}{zf} H_3
+\frac{ik}{f} H'_v -\frac{3ik}{zf} H_v=0, \nonumber
 \end{align}
where  $f=1-z^4$.
 The above differential equations can be  decoupled  to give the following equations by considering  $H'_v = p_v$,
\begin{eqnarray}
v<0:& &\quad z^2 p''_v + z (-3+2i\omega z) p'_v-(-3+k^2z^2+3i\omega z)p_v = 0, \label{ads5shear1}\\
v>0:& & \quad z^2 (1-z^4) p''_v + z (-3-z^4+2i\omega z) p'_v+(3-k^2z^2+9z^4-3i\omega z)p_v = 0. \nonumber \\ \label{ads5shear2}
\end{eqnarray}

\paragraph{Solution in $AdS_5$: $v<0$}

\ \ \\
We see that on substituting $m^2=0$, i.e. $\nu=1$ in (\ref{scalarads}), the equation reduces to (\ref{ads5shear1}). 
Therefore we can read out the solution in mixed Fourier space which satisfies the boundary condition (\ref{gen-0-0}) 
from (\ref{ads5bessel-sol}). This is given by 
\begin{align}
p^{AdS}_v & (v,k,z;t_1)=G_R^{AdS}(v,k,z;t_1) =\\
& \theta(v-t_1)2^{-\frac{1}{2}}\sqrt{\pi}
\left(\frac{k}{\sqrt{(v-t_1)^2+2(v-t_1) z}}\right)^{\frac{3}{2}} z^3 J_{-\frac{3}{2}} \left(k \sqrt{(v-t_1)^2+2(v-t_1) z} \right). \nonumber
\end{align}

\paragraph{Solution for $AdS_5$ black hole: $v>0$}
\ \ \\

We again use the  Frobenius expansion around the boundary $z=0$   to solve  (\ref{ads5shear2}). 
We obtain the following   recursion relation for $a_n^j$  for the 
the root $\Delta_+ =  3$  of the indicial equation where  $a_0^0=1$,
\begin{eqnarray} \label{vectoran1}
&& a_1^0=0, \quad a_1^1=-ia_0^0, \\
&& a_2^0 = \frac{k^2 a_0^0}{8}, \quad a_2^1=0, \quad a_2^2 = \frac{-i 5 a_1^1}{8},\nonumber \\
&& a_3^0 = 0, \quad a_3^1=\frac{k^2a_1^1-i7a_2^0}{15}, \quad  a_3^2=0, \quad a_3^3 = \frac{-i7a_2^2}{15}, \nonumber
\end{eqnarray}
and, for $n \geq 4$,
\begin{align} \label{vectoran2}
a_{n}^0 & = \frac{k^2 a_{n-2}^0+(n^2-2n-8)a_{n-4}^0}{n(n+2)},\\
a_{n}^j & = \frac{-i(2n+1) a_{n-1}^{j-1} +k^2 a_{n-2}^j+(n^2-2n-8)a_{n-4}^j}{n(n+2)}; \quad j=1,2, \dots, n-4, \nonumber \\
a_{n}^j & = \frac{-i(2n+1) a_{n-1}^{j-1} +k^2 a_{n-2}^j}{n(n+2)}; \quad j=n-3,n-2, \nonumber\\
a_{n}^j & = \frac{-i(2n+1) a_{n-1}^{j-1}}{n(n+2)}; \quad j=n-1,n. \nonumber
\end{align}
Note that  $a_n^j$ is defined in (\ref{defan}) determine the Frobenius expansion around the boundary 
for the equation (\ref{ads5shear2}).
The  thermalizing  Green function $\langle T_{t x^1} T_{t x^1} \rangle$ can be 
read out  by  by applying the matching condition,  (\ref{gen-1}), 
on $p_v$.  This is because $H_v$ can be obtained by integrating $p_v$. This just introduces a power of $z$ which
adjusts to give the right $\Delta_+$ root for the fluctuation $H_v$. Now $H_v$ is related to the vector fluctuation $h_{vx^1}$ which 
reduces to $h_{tx^1}$ at the boundary $v$ equals the boundary time $t$. Therefore 
reading out the time dependence of $p_v$ is sufficient to extract the time dependence of the 
the two point function $\langle T_{t x^1} T_{t x^1} \rangle$.

\begin{figure}
\centering
\begin{subfigure}{.5\textwidth}
  \centering
  \includegraphics[scale=0.8]{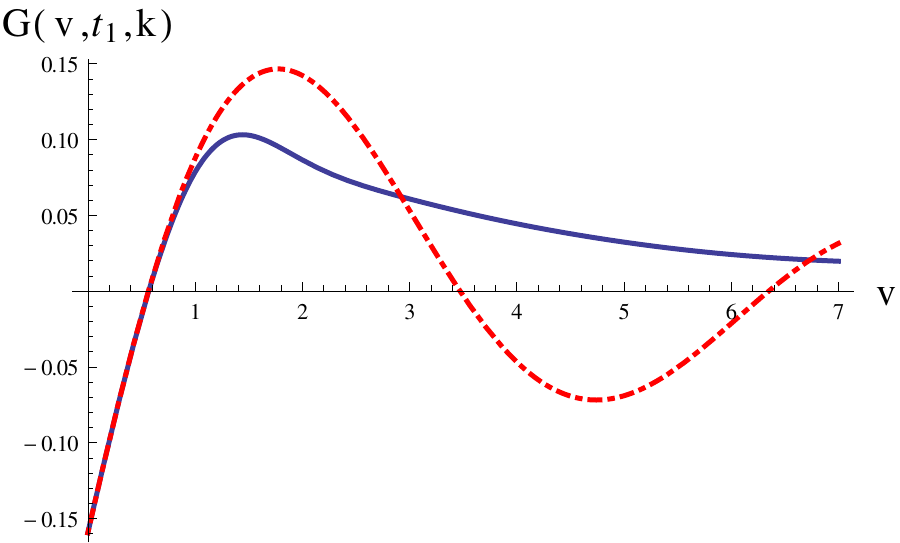}
  \caption{•}
  \label{fig-shear-a}
\end{subfigure}%
\begin{subfigure}{.5\textwidth}
  \centering
  \includegraphics[scale=0.8]{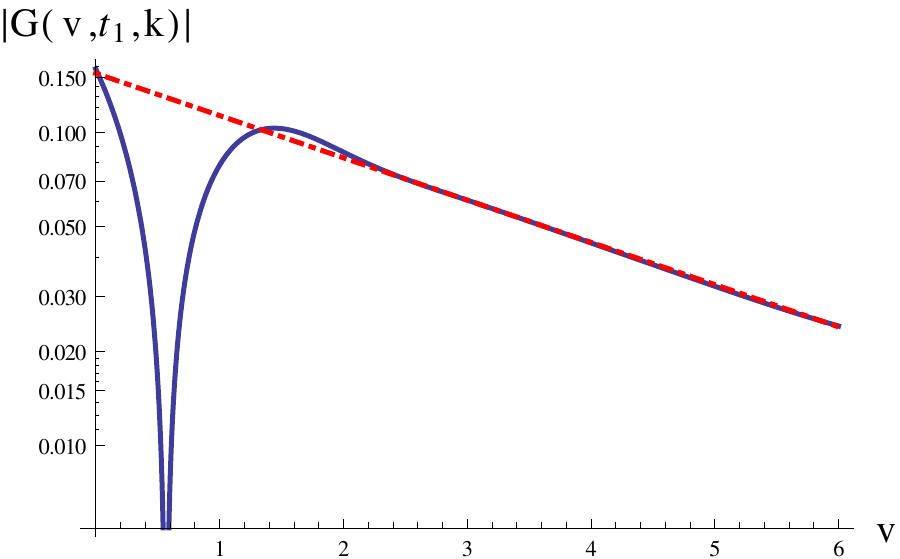}
  \caption{•}
  \label{fig-shear-b}
\end{subfigure}
\caption{The solid blue curve is the thermalizing Green function. In (a), the dot-dashed curve is the vacuum Green function. Figure (b) has the logarithmic plot of the absolute value of the Green function, plotted with its late time slope (red dot-dashed curve). Plots are for values, $t_1=-5$ and $k=1.1$.}
\label{fig-shear}
\end{figure}

\paragraph{Green function and quasinormal modes}
\ \ \\
\begin{figure}
\centering
\includegraphics[scale=1]{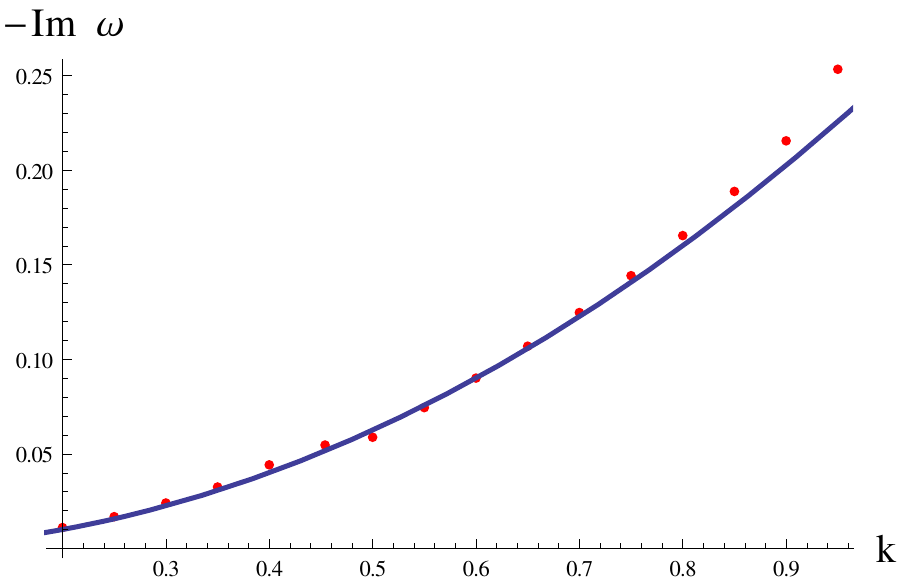}
\caption{The red dots show the value of the hydrodynamic frequency obtained from the slope of the long time behaviour of the Green function (at $t_1=-5$) for various values of $k$. The blue curve plots the expected behaviour, $-$Im $\omega = \frac{1}{4} k^2$.}
\label{fig-parab}
\end{figure}
Using  the equations (\ref{vectoran1}),(\ref{vectoran2}),(\ref{recurm}) and (\ref{recurjn}), 
with $\nu=1$, the moments $M_i$ are obtained.  Then the 
retarded two point function of ($v$,$x^1$) component of the stress energy tensor of the boundary theory, 
is evaluated  using (\ref{greenp}). We used Mathematica  to evaluate  up to 
3500 moments $C_2(\omega)$,  the results converged at this order of moments. 
The Green function is constructed to 
$O(v^{3500})$ in future time, $v$ and then we approximated it by its   $(1750|1750)$ Pad\'{e} approximant. From these moments the Green function is  evaluated for various 
values of the momentum $k$. 

Figure \ref{fig-shear-a} shows the thermalizing Green function (solid blue curve), which starts close to the 
vacuum Green function (dot-dashed red curve), and deviates away from it within one horizon time.

The  hydrodynamic pole  of the thermal two point function $\langle T_{t x^1} T_{t x^1} \rangle$ or the quasinormal mode
of the vector perturbations of the metric 
obeys the dispersion relation 
\begin{align} 
\omega = - i \frac{\eta}{s T} k^2.
\end{align}
$\eta$, $s$ and $T$ are the shear viscosity, entropy density and temperature, respectively. 
The  universal  value of $\eta/s$ for the vector mode of metric perturbation is known to be $(4 \pi)^{-1}$. 
The temperature of the  $AdS_5$ black hole is  $T=\pi^{-1}$, 
in which the radius of  $AdS$ and that of the horizon are normalized to unity. 
Therefore the dispersion relation for the hydrodynamic quasinormal frequencies is given by 
\begin{align} \label{hydroqn}
\omega = - i \frac{k^2}{4}.
\end{align}
As discussed in section \ref{longt}  the long
 time behaviour of the thermalizing Green function is dictated by the 
lowest quasinormal mode. Therefore  it is possible to extract 
the value of  the $-{\rm Im} \omega$ from 
 the slope of the logarithmic plot of the thermalizing Green function for $k<<1$ and show 
 that it obeys the dispersion relation given in (\ref{hydroqn}).

In figure \ref{fig-shear-b}, the logarithmic plot of the absolute value of the thermalizing 
Green function (solid blue curve) is plotted along with the line measuring the slope of its large time 
behaviour (dot-dashed red line). The slope of this straight line gives the value of $-$Im $\omega$. 
The plot is for fixed values of $t_1=-15$ and $k=10$. Figure \ref{fig-parab}, shows the red dots for values of $-$Im $\omega$ 
for several values of $k$ obtained from  Mathematica. 
 For small $k$, these values lie very close to the expected parabolic behaviour (solid blue curve), 
predicted by the hydrodynamic dispersion relation
(\ref{hydroqn}). 
For larger $k$, these values lie inside the parabola. This behaviour agrees with figure 14 of 
\citep{Nunez:2003eq}  which studied this quasinormal mode for large values of $k$ in the 
$AdS_5$ black hole.  

\section{Conclusions}

We have developed a recursive method to obtain the time dependent Green functions 
in the thin shell $AdS$ Vaidya background. Using the intuition developed from this method
we showed that the long time behaviour of the Green function is determined by the 
lowest quasinormal mode of the corresponding black hole.  Thus our analysis provides
an explicit realization of the general conjecture made in \cite{Horowitz:1999jd} that time scales 
in thermalization are determined by the quasinormal modes. 
We applied the method to study Green functions in  thin shell Vaidya geometries in 
$AdS_3$ and  $AdS_5$. Using this method we obtained the universal 
ratio of shear viscosity to entropy density by studying the relaxation of the 
time dependent Green function of the vector metric perturbation in the 
$AdS_5$ Vaidya shell.

The methods developed in this paper to study time dependent Green functions can be  
generalized for  fermion Green function which was obtained  by direct numerical integration
in \cite{Callebaut:2014tva}. It will be interesting to generalize this method to Vaidya  shell geometries which 
have a finite width found in \cite{Bhattacharyya:2009uu}. These geometries 
are a more accurate description of the thermalization process. The dependence of the width of the 
shell on the time dependence of the Green function will be interesting to extract. 
It is interesting to study  thermalization of Green function of higher spin fields 
to study the spin dependence in thermalization. 
It is known that a non local probe like entanglement entropy is the slowest to relax to equilibrium
\cite{Balasubramanian:2011ur}. It is will be interesting to see if higher spin fields relax faster or slower compared to scalars
and entanglement entropy. 
In this context the exact solutions to wave functions of higher spin fields found
in  the BTZ background \cite{Datta:2011za,Datta:2012gc} will prove to be useful to obtain analytical results. 
Finally it will be useful to use these lessons learned in holography to understand time dependent Green function 
in conformal  field theories  during thermalization along the lines developed in \cite{Calabrese:2006rx}. 

\acknowledgments

We thank Johanna Erdmenger, Daniel Fernandez  and Eugenio Megias-Fernandez  for useful discussions 
during the initial phase of this project. 
We thank  Sayantani Bhattacharyya, Chethan Krishnan, 
 Kalyana Rama and Bala Sathiapalan for asking questions 
that clarified our understanding of the problem.  We also thank Pallab Basu and Dileep Jatkar for 
interesting  comments. 
S.K. thanks Shouvik Datta for help with the 
graphics used in the paper. 
We thank the participants of the Bangalore area string meeting organized by the ICTS, TIFR, the string groups
of IMSc, Chennai and IIT Kanpur for  opportunities  to present the results at various stages of the project and 
the resulting discussions.  Finally J.R.D.  thanks Dieter Lust and the string group at Max Planck Institute for Physics,  Munich 
for hospitality during the initial phase of this project.

\appendix
\section{Scalar wave functions in BTZ} \label{appendix-btz}

In this appendix we derive the scalar wave functions in the BTZ geometry as 
an expansion around the boundary by transforming the known wave functions which are written as 
expansions around the horizon. 
Consider the wave functions given in (\ref{btzsol-horizon}) which admit a natural expansion
near the horizon. We can use the following  transformation property of the Hypergeometric function to 
write it as an expansion around the boundary,
\begin{align}\label{conformula}
F(a,b,c,z) = & \frac{\Gamma(c)\Gamma(c-a-b)}{\Gamma(c-a) \Gamma(c-b)} F(a,b, a+b-c+1, 1-z)\\
& + (1-z)^{c-a-b} \frac{\Gamma(c) \Gamma(a+b-c)}{\Gamma(a) \Gamma(b)} F(c-a,c-b,c-a-b+1,1-z). \nonumber
\end{align}
Substituting (\ref{conformula}) in  (\ref{btzsol-horizon})  we obtain
\begin{eqnarray} \label{btz-bdy1}
\Phi_{\omega, k }^{(1)} (z, v, \phi) && = e^{ - i \omega v + i k \phi} (1+ z)^{-i \omega} \times\\
&& \Bigg[ B_1(\nu,\omega,k) z^{\Delta_-} F \left(\frac{1}{2} (\Delta_- - i(\omega - k)),\frac{1}{2}(\Delta_- - i (\omega + k)),\Delta_-, z^2\right) \nonumber\\ 
+ && B_2(\nu,\omega,k) z^{\Delta_+} F \left(\frac{1}{2} (\Delta_+ - i(\omega + k)),\frac{1}{2}(\Delta_+ - i (\omega - k)),\Delta_+, z^2\right) \Bigg], \nonumber
\end{eqnarray}
\begin{eqnarray} \label{btz-bdy2}
 \Phi_{\omega, k }^{(2)} (z, v, \phi) && = e^{ - i \omega v + i k \phi} (1- z)^{i \omega} \times\\
 && \Bigg[B_3(\nu,\omega,k) z^{\Delta_-} F \left(\frac{1}{2} (\Delta_- + i (\omega + k)),\frac{1}{2} (\Delta_- + i(\omega - k)),\Delta_-, z^2 \right) \nonumber\\
+ && B_4(\nu,\omega,k) z^{\Delta_+} F \left(\frac{1}{2} (\Delta_+ + i (\omega - k)),\frac{1}{2} (\Delta_+ + i(\omega + k)),\Delta_+, z^2 \right) \Bigg].\nonumber
\end{eqnarray}
The coefficients $B_i$, $i=1, \dots 4$, are made up of gamma functions obtained from
(\ref{conformula}), and do not depend on $z$. 
The hypergeometric functions of (\ref{btz-bdy2}) can further be related 
to the hypergeometric functions of (\ref{btz-bdy1}), by the transformation, 
\begin{align}
F(a,b,c,z) = (1-z)^{c-a-b} F(c-a,c-b,c,z).
\end{align}
Thus, the only two independent near boundary solutions are,
\begin{eqnarray}
 \Phi_{\omega, k}^{BTZ (-)} (z, v, \phi) & =& e^{ - i \omega v + i k \phi} z^{\Delta_-} 
 ( 1+ z)^{i \omega}  \times \\ \nonumber
& & F \left(\frac{1}{2} (\Delta_- - i(\omega - k)),\frac{1}{2}(\Delta_- - i (\omega + k)),\Delta_-,  z^2\right),
\\ \nonumber
  \Phi_{\omega, k}^{BTZ (+)} (z, v, \phi) & =& e^{ - i \omega v + i k \phi} z^{\Delta_+} 
  ( 1+ z)^{i \omega}  \times \\ \nonumber
& &  F \left(\frac{1}{2} (\Delta_+ - i(\omega + k)),\frac{1}{2}(\Delta_+ - i (\omega - k)),\Delta_+, z^2\right).
\end{eqnarray}

\section{Thermal Green function in BTZ} \label{app-thermal}
The thermal Green function is obtained by taking the Fourier transform 
with respect to $\omega$ of the following thermal Green function in  Fourier space  \cite{Balasubramanian:2012tu},
\begin{align}
\begin{split}
G_R^{\text{thermal}}(k,\omega) = & \frac{1}{2 \sin(\pi \Delta_+) (\Gamma(\Delta_+))^2} \left| \Gamma \left( \frac{1}{2}(\Delta_++i(\omega+k)) \right)\Gamma \left( \frac{1}{2}(\Delta_++i(\omega-k)) \right) \right|^2\\
& \times \left( \cos(\pi \Delta_+) \cosh(\pi \omega) - \cosh(\pi k) -i \sin(\pi \Delta_+) \sinh(\pi \omega) \right).
\end{split}
\end{align}
The Fourier transform is taken by performing  integration over $\int_{-\infty}^{\infty} d\omega e^{-i\omega(v-t_1)}$ and
completing the contour in the lower half $\omega$-plane.  It can be shown that the integral vanishes 
over the arc of sufficiently large radius. 
The result is a sum over the residues, which are evaluated at the poles of the gamma functions in the lower half $\omega$-plane,
\begin{align}\label{thermal}
G_R^{\text{thermal}}(v-t_1,\omega) = & \frac{1}{2 \sin(\pi \Delta_+) (\Gamma(\Delta_+))^2}\\
& \times \sum_n \sum_{a=\pm 1} \frac{(-1)^n}{n!} e^{-i \omega_n^a (v-t_1)} \Gamma \left(\frac{1}{2} (\Delta_+ + i(\omega_n^a-a k)) \right) \nonumber\\
& \times \Gamma \left(\frac{1}{2} (\Delta_+ + i(\omega_n^a+a k)) \right)\Gamma \left(\frac{1}{2} (\Delta_+ - i(\omega_n^a+a k)) \right) \nonumber\\
& \times \left[\cos(\pi \Delta_+) \cosh(\pi \omega_n^a)-\cosh(\pi k) -i \sin(\pi \Delta_+) \sinh(\pi \omega_n^a) \right].\nonumber
\end{align}
These poles, $\omega(n,a)$, are the quasinormal modes of the BTZ black hole,
\begin{align}
\omega_n^a = \, & a k -i(\Delta_+ + 2n), \qquad a = \pm 1.
\end{align}

\section{ Mixed Fourier transform of Green function in  $AdS$} \label{appendix_ads5}

In this appendix we solve equations of the form (\ref{ads5shear2}) with the analog of the  boundary conditions in (\ref{gen-0-0}) 
first in the frequency-momentum space.  We then perform the partial Fourier transform in the frequency space. 
This results in the Green function in $AdS$ which satisfies the boundary condition of (\ref{gen-0-0}). 
Consider (\ref{ads5shear2}), with a more general mass term,
\begin{align} \label{scalarads}
\quad z^2 p''_v + z (-3+2i\omega z) p'_v-(-3+m^2+k^2z^2+3i\omega z)p_v = 0,
\end{align}
this equation has the following Fourier space solution, where $\nu = \sqrt{1+m^2}$,
\begin{align} \label{adsbessel}
p_v(k,\omega,z) = e^{-i\omega z} z^2 \left(A(\omega) J_\nu \left(z \sqrt{\omega^2-k^2}\right)+B(\omega) J_{-\nu} \left(z \sqrt{\omega^2-k^2}\right)\right).
\end{align}
Using the property of the Bessel function for small argument, the following near boundary behaviour of the solution is obtained,
\begin{align}
p_v(k,\omega,z) \xrightarrow{z \to 0} z^2 \left(A(\omega) \left(\frac{z \sqrt{\omega^2-k^2}}{2}\right)^\nu \frac{1}{\Gamma(1+\nu)}+B(\omega) \left(\frac{z \sqrt{\omega^2-k^2}}{2}\right)^{-\nu} \frac{1}{\Gamma(1-\nu)}\right).
\end{align}
The delta function boundary condition in (\ref{gen-0-0})  in the  time domain requires that near the boundary 
the wave function in  Fourier space solution goes to
\begin{equation}
p_v(k,\omega,z) \xrightarrow{z \to 0} \frac{1}{2\pi}  \cdot z^{\Delta_-}.
\end{equation}
This determines  $B(\omega)$ to be 
\begin{align}
B(\omega) = \frac{\Gamma(1-\nu)}{2\pi} \left(\frac{\sqrt{\omega^2-k^2}}{2}\right)^\nu,
\end{align}
where, $\Delta_\pm = 2\pm \nu$.

We now use the following asymptotic behaviour of  Bessel functions to obtain the behaviour of the 
solution near the origin of $AdS$, 
\begin{align}
J_\nu(x) & \xrightarrow{x \to \infty} \sqrt{\frac{2}{\pi x}} \cos\left(x-\frac{\nu \pi}{2}-\frac{\pi}{4}\right),\\ 
J_{-\nu}(x) & \xrightarrow{x \to \infty} \sqrt{\frac{2}{\pi x}}\left[\cos (\pi \nu) \cos\left(x-\frac{\nu \pi}{2}-\frac{\pi}{4}\right)-\sin (\pi \nu) \sin \left( x-\frac{\nu \pi }{2} - \frac{\pi}{4} \right) \right].\nonumber
\end{align}
As $z \to \infty$, the solution,
\begin{align}
\begin{split}
p_v(k,\omega,z) \longrightarrow e^{-i\omega z} z^2 \sqrt{\frac{2}{\pi z}} (\omega^2-k^2)^{-\frac{1}{4}} & \Big[ \left(A(\omega) + B(\omega) \cos(\pi \nu) \right) \cos\left(x-\frac{\nu \pi}{2}-\frac{\pi}{4}\right)\\
& - B(\omega) \sin(\pi \nu) \sin \left(x - \frac{\pi \nu}{2} -\frac{\pi}{4}\right) \Big],
\end{split}
\end{align}
where, $x=z \sqrt{\omega^2-k^2}$. 
The solution near the origin has both ingoing and outgoing 
behaviour due to the presence of cosine and sine. 
In order to impose ingoing boundary condition at origin, the solution is 
rewritten as a linear combination of ingoing and outgoing solutions,
\begin{align}
p_v(k,\omega,z) = C(\omega) e^{-i(x-\frac{\nu \pi}{2}-\frac{\pi}{4})}+D(\omega) e^{i(x-\frac{\nu \pi}{2}-\frac{\pi}{4})},
\end{align}
where $C$ and $D$ in terms of $A$ and $B$, are
\begin{align}
C(\omega) & = \frac{1}{2} \left( A(\omega)+B(\omega) e^{-i\pi \nu} \right),\\
D(\omega) & = \frac{1}{2} \left( A(\omega)+B(\omega) e^{i\pi \nu} \right).
\end{align}
The ingoing boundary condition at origin, means that $D(\omega)=0$, hence,
\begin{align}
A(\omega) = - B(\omega) e^{i \pi \nu}.
\end{align}
Substituting this in (\ref{adsbessel}), we obtain,
\begin{align}
p_v(k,\omega,z) & = B(\omega) e^{-i \omega z} z^2 \left[-e^{i\pi \nu} J_\nu(x)+J_{-\nu}(x)\right]\\
& = -i \sin(\pi \nu)e^{i\pi\nu} B(\omega) e^{-i \omega z} z  H_{-\nu}^{(2)}(x).\nonumber
\end{align}
The Fourier transform of the above equation with respect to $\omega$, gives
\begin{align}
G_F^{AdS} & (v,k,z) =\\
& -i e^{i\pi\nu} \frac{1}{2\Gamma(\nu)} z^2 \int_{-\infty}^\infty d\omega \, e^{-i\omega(v+z)} \left( \frac{\sqrt{\omega^2-k^2}}{2} \right)^\nu H_{-\nu}^{(2)}(z\sqrt{\omega^2-k^2}),\nonumber
\end{align}
using the integral representation of $H_{-\nu}^{(2)}(x)$,
\begin{align}
H^{(2)}_{-\nu}(x)=\frac{i}{\pi} e^{-\frac{i\pi\nu}{2}} x^{-\nu} \int_0^\infty dy \, \exp\left[ -\frac{i}{2}\left(y+\frac{x^2}{y}\right) \right] y^{\nu-1},
\end{align}
\begin{align}
G_F^{AdS} & (v,k,z) = \; \frac{ e^{\frac{i\pi\nu}{2}}}{2^\nu 2\pi \Gamma(\nu)} z^{2-\nu} \int_0^\infty dy \, y^{\nu-1} \exp\left[-\frac{i}{2}\left(y-\frac{z^2k^2}{y}\right)\right]\\
& \times \int_{-\infty}^\infty d\omega \, \exp\left[-i\left(\frac{\omega^2z^2}{2y}+\omega(v+z)\right)\right] \nonumber\\
= & \; \frac{e^{-i\frac{\pi}{4}} e^{\frac{i\pi\nu}{2}}}{2^\nu 2 \pi \Gamma(\nu)} \sqrt{2\pi} z^{1-\nu} \int_0^\infty dy \, y^{\nu-\frac{1}{2}} \exp\left[\frac{i}{2}\left(y\left(\frac{2v}{z}+\left(\frac{v}{z}\right)^2\right)+\frac{z^2k^2}{y}\right)\right] \nonumber\\
=& \; \frac{e^{-i\frac{\pi}{4}} e^{\frac{i\pi\nu}{2}}}{2^\nu 2\pi \Gamma(\nu)} \sqrt{2\pi}z^{1-\nu} 2 \left(\frac{k z^2}{\sqrt{v^2+2v z}}\right)^{\nu+\frac{1}{2}} K_{-\nu-\frac{1}{2}}(-i k \sqrt{v^2+2 v z}) \nonumber\\
=& \; \frac{1} {2^\nu \Gamma(\nu) } \frac{\sqrt{2\pi}}{2}z^{2+\nu} \left(\frac{k}{\sqrt{v^2+2v z}}\right)^{\nu+\frac{1}{2}} H^{(1)}_{-\nu-\frac{1}{2}}(k \sqrt{v^2+2 v z}). \nonumber
\end{align}
In the last step, $K_\lambda(x)=\frac{i\pi}{2} e^{\frac{i\pi \lambda}{2}} H_\lambda^{(1)}(i x)$, has been used. 
The retarded two point function is obtained from the Feynman two point function using the relation
\begin{align}
G_R(t,t')=\theta(t-t') \left(G_F(t,t')+G^*_F(t,t')\right).
\end{align}
The following properties of the Bessel functions,
\begin{eqnarray}
&& H^{(1)}_\lambda(x)^*=H^{(2)}_\lambda(x),\\
&& H^{(1)}_\lambda+H^{(2)}_\lambda = i \csc(\pi \lambda) (e^{-i\pi \lambda}-e^{i\pi\lambda})J_\lambda(x), \nonumber
\end{eqnarray} 
are used in last two equations,  to obtain the retarded Green  function which is given by 
\begin{align} \label{ads5bessel-sol}
G_R^{AdS}(v-t_1,k,z)  = C& \theta(v-t_1) z^{2+\nu} \left(\frac{k}{\sqrt{(v-t_1)^2+2(v-t_1) z}}\right)^{\nu+\frac{1}{2}} \\ 
& \times J_{-\nu-\frac{1}{2}}\left(k \sqrt{(v-t_1)^2+2 (v-t_1) z}\right),\nonumber
\end{align}
where, $C=\frac{2^{\frac{1}{2}-\nu}\sqrt{\pi}}{\Gamma(\nu)}$. Though we started out by assuming 
$\nu$ is not an integer and the two independent solutions are given by (\ref{adsbessel}), the discussion can be 
generalized to the case when $\nu$ is an integer, leading to the same final result given in (\ref{ads5bessel-sol}). 
A simple way to see  this is that in the final result the order of the Bessel function is fractional. 

\section{Recursion relation for retarded Green functions in $AdS$ }

In this appendix for completeness we obtain the recursion relation for the mixed Fourier space retarded Green function 
in $AdS_{d+1}$ at $v=0$. 
This is useful to implement the recursive algorithim to obtain the thermalizing Green function 
in thin shell Vaidya geometries. 
The solution for a scalar field in mixed Fourier space in AdS$_{d+1}$, 
has the following dependence on the Bessel function,
\begin{align}
\Phi^{AdS}(v=0,k,z;t_1) = z^{\Delta_+} R(z).
\end{align}
Here we have assumed $t_1<0$. 
\begin{align} \label{bessel-gen}
R(z) = C \left(\frac{k}{\sqrt{t_1^2-2 t_1 z}} \right)^{\nu+\frac{1}{2}} J_{-\nu-\frac{1}{2}} \left(k \sqrt{t_1^2-2 t_1 z} \right),
\end{align} 
where, from (\ref{ads3-5}) and (\ref{ads5bessel-sol}), we see that, $\Delta_+=d/2+\nu$.
The differential equation satisfied by $R(z)$ is,
\begin{align}
(t_1-2z) R''(z) + 2(\lambda-1)R'(z)+k^2 t_1 R(z)=0.
\end{align}
Substituting the ansatz, $R(z) = \sum_{n=0}^\infty \tilde{J}_n z^n$ in the above equation, the following recursion relation for $\tilde{J}_n$ is obtained,
\begin{align} \label{recurjn}
\tilde{J}_n = \frac{2(n-1)(n-1-\lambda)\tilde{J}_{n-1}-k^2 t_1 \tilde{J}_{n-2}}{t_1 n(n-1)}; \qquad \text{for} \, n \geq 2.
\end{align}
To get the solution of (\ref{bessel-gen}) from the above recursion relation, $\tilde{J}_0$ and $\tilde{J}_1$ are chosen to be,
\begin{align}
\tilde{J}_0 & = C \left(\frac{k}{ |t_1|}\right)^{\nu+\frac{1}{2}} J_{-\nu-\frac{1}{2}}(|k||t_1|),\\
\tilde{J}_1 & = C \left(\frac{k}{ |t_1|}\right)^{\nu+\frac{1}{2}} \, k \, J_{-\nu-\frac{3}{2}}(|k||t_1|).
\end{align}
Thus we obtain the Frobenius series expansion of AdS scalar field solution, around $z=0$,
\begin{align}
\Phi^{AdS}(v=0,k,z;t_1) = z^{\Delta_+} \sum_{n=0}^\infty \tilde{J}_n z^n.
\end{align}

\section{Details of Mathematica files}
The recursion method of section \ref{sec-gen} is used to numerically construct the Green function 
using Mathematica for the three cases considered in this paper. The files for each case is attached.
\begin{enumerate}
\item \underline{ads3-green.nb}: The Green function for scalar field in $AdS_3$ Vaidya is 
numerically constructed to order 56 in moments and figures \ref{fig1} and \ref{fig2} are obtained.
\item \underline{shear-green.nb}: The Green function for shear metric perturbations $h_{x^1 x^2} \neq 0$ in $AdS_5$ 
Vaidya is numerically constructed to order 200 in moments and figure  \ref{fig-scalar} is obtained.
\item \underline{vector-green.nb}: The Green function for metric perturbations $h_{v x^1},h_{x^1 x^3} \neq 0$ in $AdS_5$ 
Vaidya is numerically constructed to order 3500 in moments and figure \ref{fig-shear} is obtained. 
\end{enumerate}

\bibliography{therm}
\bibliographystyle{JHEP}

\end{document}